\documentclass[twocolumn]{aastex631}

\usepackage{graphicx}% Include figure files
\usepackage{dcolumn}% Align table columns on decimal point
\usepackage{bm}% bold math
\usepackage{amsmath}
\usepackage{overpic}
\usepackage{booktabs}%
\usepackage{makecell}
\usepackage{enumerate}
\usepackage[figuresright]{rotating}
\makeatletter

\newcommand{\Rmnum}[1]{\expandafter\@slowromancap\romannumeral #1@}
\makeatother
\usepackage{hyperref}
\usepackage{xcolor}


\setwatermarkfontsize{1.5in}

\shortauthors{Tan ET AL.}

\begin{document}

\title{Search for Type IL Gamma-ray Bursts: Criterion, Results, Verification and Physical Implication}

\correspondingauthor{Shao-Lin Xiong}
\email{xiongsl@ihep.ac.cn}

\author{Wen-Jun Tan}
\affil{State Key Laboratory of Particle Astrophysics, Institute of High Energy Physics, Chinese Academy of Sciences, Beijing 100049, China}
\affil{University of Chinese Academy of Sciences, Chinese Academy of Sciences, Beijing 100049, China}

\author{Chen-Wei Wang}
\affil{State Key Laboratory of Particle Astrophysics, Institute of High Energy Physics, Chinese Academy of Sciences, Beijing 100049, China}
\affil{University of Chinese Academy of Sciences, Chinese Academy of Sciences, Beijing 100049, China}

\author{Peng Zhang}
\affil{State Key Laboratory of Particle Astrophysics, Institute of High Energy Physics, Chinese Academy of Sciences, Beijing 100049, China}
\affil{College of Electronic and Information Engineering, Tongji University, Shanghai 201804, China}

\author{Wang-Chen Xue}
\affil{State Key Laboratory of Particle Astrophysics, Institute of High Energy Physics, Chinese Academy of Sciences, Beijing 100049, China}
\affil{University of Chinese Academy of Sciences, Chinese Academy of Sciences, Beijing 100049, China}

\author{Shao-Lin Xiong*}
\affil{State Key Laboratory of Particle Astrophysics, Institute of High Energy Physics, Chinese Academy of Sciences, Beijing 100049, China}

\author{Bo-bing Wu}
\affil{State Key Laboratory of Particle Astrophysics, Institute of High Energy Physics, Chinese Academy of Sciences, Beijing 100049, China}

\author{Jia-Cong Liu}
\affil{State Key Laboratory of Particle Astrophysics, Institute of High Energy Physics, Chinese Academy of Sciences, Beijing 100049, China}
\affil{University of Chinese Academy of Sciences, Chinese Academy of Sciences, Beijing 100049, China}

\author{Yue Wang}
\affil{State Key Laboratory of Particle Astrophysics, Institute of High Energy Physics, Chinese Academy of Sciences, Beijing 100049, China}
\affil{University of Chinese Academy of Sciences, Chinese Academy of Sciences, Beijing 100049, China}

\author{Sheng-Lun Xie}
\affil{Institute of Astrophysics, Central China Normal University, Wuhan 430079, China}
\affil{State Key Laboratory of Particle Astrophysics, Institute of High Energy Physics, Chinese Academy of Sciences, Beijing 100049, China}

\author{Zheng-Hang Yu}
\affil{State Key Laboratory of Particle Astrophysics, Institute of High Energy Physics, Chinese Academy of Sciences, Beijing 100049, China}
\affil{University of Chinese Academy of Sciences, Chinese Academy of Sciences, Beijing 100049, China}

\author{Jin-Peng Zhang}
\affil{State Key Laboratory of Particle Astrophysics, Institute of High Energy Physics, Chinese Academy of Sciences, Beijing 100049, China}
\affil{University of Chinese Academy of Sciences, Chinese Academy of Sciences, Beijing 100049, China}

\author{Wen-Long Zhang}
\affil{State Key Laboratory of Particle Astrophysics, Institute of High Energy Physics, Chinese Academy of Sciences, Beijing 100049, China}
\affil{School of Physics and Physical Engineering, Qufu Normal University, Qufu, Shandong 273165, China}

\author{Yan-Qiu Zhang}
\affil{State Key Laboratory of Particle Astrophysics, Institute of High Energy Physics, Chinese Academy of Sciences, Beijing 100049, China}
\affil{University of Chinese Academy of Sciences, Chinese Academy of Sciences, Beijing 100049, China}

\author{Chao Zheng}
\affil{State Key Laboratory of Particle Astrophysics, Institute of High Energy Physics, Chinese Academy of Sciences, Beijing 100049, China}
\affil{University of Chinese Academy of Sciences, Chinese Academy of Sciences, Beijing 100049, China}

\begin{abstract}
As an interesting subclass of gamma-ray burst (GRB), Type IL GRB (such as GRB 211211A and GRB 230307A) features a long-duration prompt emission but originating from compact binary merger. The ``long duration” emisison of Type IL GRB are dominately composed of the main burst, rather than the extended emission,  
differentiating them from the traditional ``long-short” GRB (e.g., GRB 060614). 
Previous study has reported several Type IL GRBs by visual inspection of their light curves. In this work, we established a detailed criterion to identify Type IL GRBs by light curve, and then systematically searched the archival \textit{Fermi}/GBM data with this criterion, resulting in a sample of 5 type IL GRBs from January 1, 2014 to January 1, 2024, i.e. GRB 230307A, GRB 211211A, GRB 200914A, GRB 200311A and GRB 170228A. 
Apart from the light curve pattern, we find that the temporal and spectral properties of these 5 GRBs also support this classification. Interestingly, we find that the energy ratio between extended emission and main emission is almost constant ($\sim0.7$, with small scattering) for these GRBs, which have strong implication on the mechanism of Type IL burst. We discuss theoretical models to interpret the progenitor, central engine, and extended emission of these Type IL bursts. 

\end{abstract}


\section{Introduction}

    As one of the most violent explosive events in the universe, Gamma-ray bursts (GRBs) have been studied intensively since the discovery in 1960s. In 1981, GRBs were classified as long bursts and short bursts in KONUS experiments \citep{twoclassGRB}. In 1993, the most well-known classification criterion for GRBs was proposed \citep{T90classification}, that is the T$_{90}$ (time interval with accumulated GRB photon counts increase from 5\% to 95\% of the full burst) was found to exhibit a bimodal distribution with a separation of about 2 seconds. 
    This classification divides GRBs into ``long” bursts and ``short” bursts. 
    
    Since the duration usually reflects the activity time of the central engine of the GRB; its bimodal distribution also implies the existence of two distinct categories of central engines, which further corresponds to two different origin models for GRBs, namely Type I (usually short) and Type II (usually long) GRBs \citep{Zhang2006type}. Type I bursts are produced by the merger of compact binaries consisting of binary neutron stars (BNS) or neutron star-black hole system (NS-BH) \citep{Blinnikovmerger1984,Paczynskimerger1986,1Meszarosmerger1992,Limerger1998}, while Type II bursts are related to the core collapse of supermassive stars \citep{Woosleycollapsar1993,Paczycollapsar1998,Woosleycollapsar2006}.

    Apart from the main emission, there are two components in the prompt emission which usually affect the calculation of the duration: precursor and extended emission (EE).
    The precursor generally refers to a brief period of weaker emission occurring before the main emission, followed by a quiescent period. 
    Some studies have indicated that precursors may contain quasi-thermal radiation components \citep{Wangpre2020}, suggesting the possibility of different radiation mechanisms compared to the main emission. This provides clues for the study of the progenitor stars of GRBs \citep{Zhongpre2019}. In practice, many searches for precursors have been conducted. The precursors were first identified in long GRBs \citep[e.g.][]{Lazzatipre2005,Burlonpre2008,Burlonpre2009,ZBB0625B2018}, and subsequently, they were also discovered in short GRBs \citep[e.g.][]{Trojapre2010,Minaevpre2017}. The common approach involves using the Bayesian Block method and background estimation to search for precursors \citep{Coppinpre2020}. The search results reveal that only a small fraction of GRBs have precursors, with the precursor and quiescent period between precurosr and main burst for long GRBs generally longer than that of short GRBs. Some GRBs even show the presence of multiple precursors \citep{Coppinpre2020}. 

    Extended emission generally refers to a prolonged period of weak radiation that occurs after the main emission. Typically, the extended emission of short GRBs receives more attention \citep{BarthelmyshortEE2005,NorrisshortEE2006,PerleyshortEE2009}. The mechanism of the production of extended emission remains an open question, with several potential models including fallback accretion or later reactivity of the central magnetar \citep{Metzger08,Gibson17,Bucciantini11}. Similarly to precursors, searches of extended emission are usually based on the Bayesian Block method and signal-to-noise ratio (SNR) \citep{2KanekoEE2015,ZhangEE2020}.

    In addition to duration, there are significant differences in the temporal and spectral characteristics of these two types of GRBs. For instance, Type I GRBs exhibit a harder energy spectrum \citep{KouveliotouHR1993}, shorter minimum variability timescales (MVT) \citep{GolkhouMVT2015}, and smaller spectral lags \citep{NorrisshortEE2006,GehrelsLAG2006} than that of Type II GRBs. Moreover, empirical correlations between various observational characteristics can serve as tools to classify GRBs, such as the Amati relation \citep{Amati2002}. In particular, the Amati relation has been utilized as a key indicator for classification of type I or type II GRBs. Overall, with the increasing observational evidence and the establishment of empirical relationships, the classification of GRBs seems to be successful for most cases.

    As the sample of the detections expands, special events emerge. Some GRBs have long durations but exhibit some properties mentioned above that fall within Type I GRB region. Even some of them (e.g. GRB 211211A, GRB 230307A) are associated with kilonova \citep{Rastinejad11Akilo,Levan07A}. There are more and more such events, leading to the colloquial designation of these GRBs as ``long" short bursts. Typical examples of ``long" short bursts include GRB 060614 \citep{GehrelsLAG2006,2006NaturGal}, GRB 211227A \citep{HouJun_211227}, and GRB 211211A \citep{YJ_GRB211211A_nature}. In fact, the light curves of GRB 060614 and GRB 211227A are quite similar, both featuring a short hard peak with a duration of approximately 5\,s, followed by a soft and prolonged extended emission lasting nearly 100\,s, which accounts for their long duration; therefore, such GRBs are also called sGRBEE. In contrast, GRB 211211A and GRB 230307A exhibit different light curve shapes from the previous two GRBs: first, they have a precursor \citep{XS_GRB211211A_QPO}; second, their main emission durations exceed 10 seconds \citep{wangtype1L}. This indicates that the long timescale of these two GRBs is not primarily due to the contribution of soft extended emission (though it also contributes some); rather, it is attributed to the hard and long main emission ($ \sim $ 10 seconds), thus distinguishing them from the typical ``long" short bursts.

    By identifying the different light curve pattern from the traditional ``long” short bursts, the concept of Type IL GRB was introduced \citep{wangtype1L}. The light curve pattern of Type IL GRB exhibits the following characteristics: short precursor, short quiescent period (that is, the interval between the precursor and the main emission), long main emission, extended emission, and a dip structure between the main emission and the extended emission. Although these GRBs have a long duration, their observational characteristics fall within Type I GRB region; hence, they are classified as a subclass of Type I GRB.

    To validate the universality of the concept of Type IL GRB and to expand Type IL GRB sample, \cite{wangtype1L} conducted a preliminary search in the \textit{Fermi}/GBM burst catalog and found GRB 170228A. Its light curve pattern and observational properties are similar to GRB 230207A and GRB 211211A, supporting its Type IL nature. However, the search in \cite{wangtype1L} solely relied on the visual identification of the precursor and extended emission and is very time-consuming.

    In this work, we establish a criterion for automatically identifying Type IL GRB based on their light curve patterns. Using the Bayesian Block method \citep{ScargleBBlock2013}, we propose stringent criteria and search methodologies to determine the presence of precursors, extended emissions, and dip structures, ultimately allowing us to determine whether a specific GRB should be classified as a Type IL GRB. We identified 5 GRBs whose light curve patterns align with the characteristics of Type IL GRBs. Upon examining their observational properties, we found that these 5 GRBs indeed fall within Type I GRB category, thereby further validating the classification of Type IL GRB as a subclass of Type I GRB. We also obtain a comprehensive sample of Type IL GRBs collected over the past decade.
    Furthermore, we also search for thermal components in the precursors of Type IL GRBs and briefly discuss theoretical models related to the progenitors, central engines, and extended radiation that could produce Type IL GRBs.

    This paper is organized as follows. Data selection required for establishing the criterion is described in section \ref{section2}. The process of criterion establishment is presented in section \ref{section3}. The results of searching based on the criterion and the verification of Type IL GRBs are given in section \ref{section4}, discussions are given in section \ref{section5}. Summary is given in section \ref{section6}. 
    
    All parameter errors in this work are for the 68\% confidence level if not otherwise stated.
    
\section{Instrument and Data selection }\label{section2}

    The Gamma-ray Burst Monitor (GBM) is one of the two instruments onboard the \textit{Fermi} Gamma-ray Space Telescope \citep{GBM_meegan_09,GBM_Bissaldi_09}, which consists of 14 detectors with different orientations: 12 Sodium Iodide (NaI) scintillation detectors (labeled from n0 to nb) covering the energy range of about 8-1000 keV, and 2 Bismuth Germanate (BGO) scintillation detectors (b0 and b1) covering the energy range of about 0.2-40 MeV.
 
    From the \textit{Fermi GBM Burst Catalog} 
    \footnote{\href{https://heasarc.gsfc.nasa.gov/W3Browse/fermi/fermigbrst.html}{https://heasarc.gsfc.nasa.gov/W3Browse/fermi/fermigbrst.html}}, 
    we selected all long GRBs with $T_{90}$ greater than 4\,s from January 1, 2014 to January 1, 2024 since the daily data of GBM were intermittent before 2013 on their website. The reason why those bursts with duration exceeding 4 seconds were chosen is because the GRBs we want to investigate generally have a long duration, for example, the main emission of GRB 211211A lasts about 10\,s. With this filter, we obtain a total dataset of about 1800 GRBs. We note that there were about 400 short GRBs in the past decade in this catalog. We use the trigger time and $T_{90}$ provided in the catalog for the subsequent criterion establishment process.

\section{Search Criterion for Type IL GRB}\label{section3}

    As mentioned in \cite{wangtype1L}, Type IL GRBs feature a unified burst pattern:
    \begin{enumerate}[(1)]
    \item there must be an evident precursor;
    \item the duration of quiescent period is defined as the waiting time, and the waiting time $T_{wt}$ should be relatively short;
    \item a relatively long $T_{90}$ of the prompt emission, including precursor, main emission and extended emission;
    \item there must be an extended emission, which is separated from the main emission by a dip-like structure.
    \end{enumerate}
    Therefore, two aspects are considered in the process of criterion establishment: one is the precursor and quiescent period, and the other is the extended emission and the dip. The standard for the long main emission is guaranteed by the filter of $T_{90}$ greater than 4\,s in the dataset.

\begin{figure*}[http]
\centering
\includegraphics[width=0.8\textwidth]{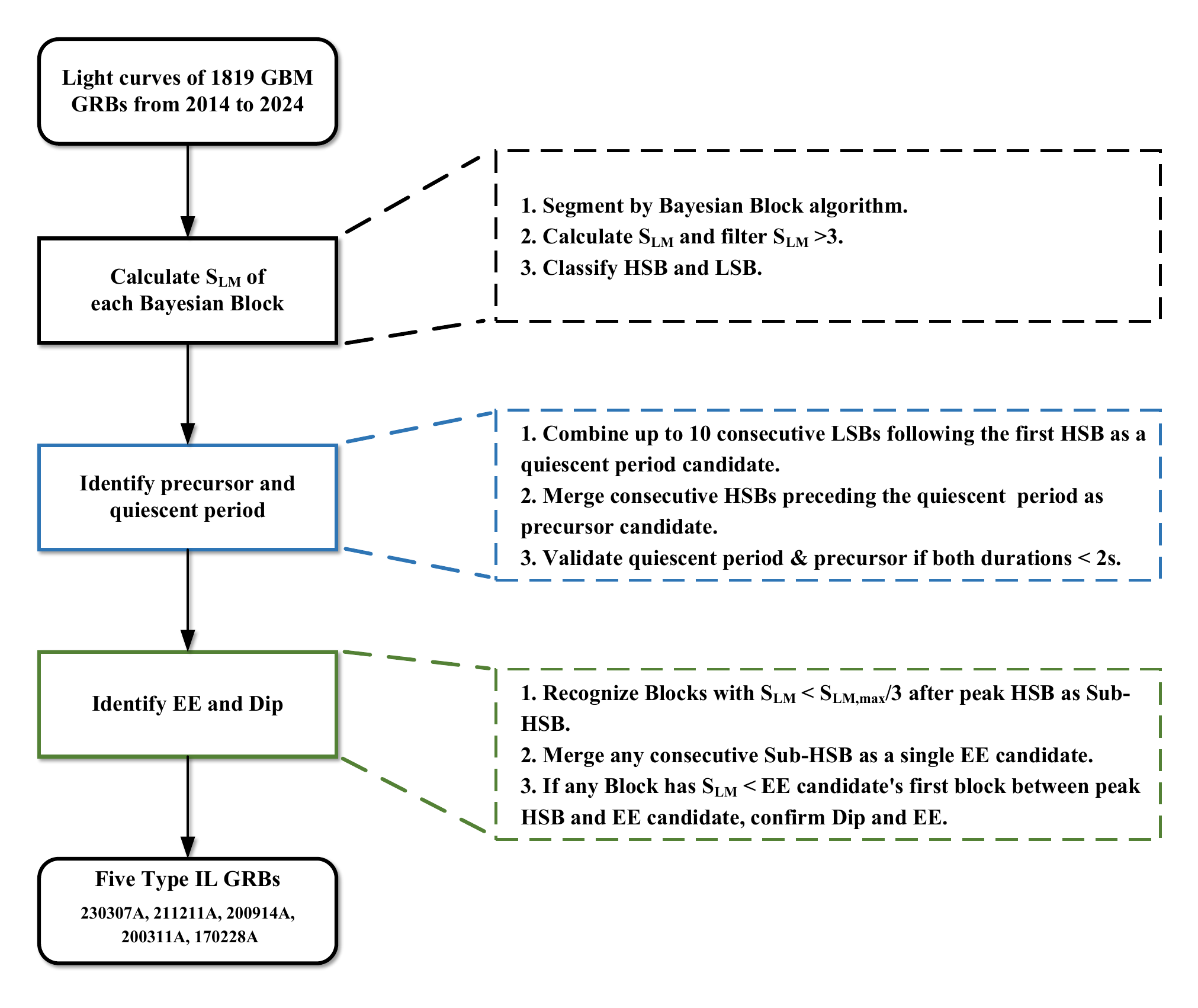}
\caption{\noindent\textbf{Flowchart of the entire criterion.}}
\label{fig:process}
\end{figure*}

     The flowchart of the criterion is shown in Fig.\ref{fig:process}. We start by extracting the total light curve of the three NaI detectors with the strongest signal from the time range T$_{0}$-5\,s to T$_{0}$+2*$T_{90}$, with a time resolution of 0.2\,s. Choosing 5\,s before T$_{0}$ is based on the consideration that the background should not have too much influence on the burst. If a GRB's precursor appears more than 5\,s before T$_{0}$, then its waiting time is too long, and this GRB should not be considered. The choice of resolution of 0.2\,s is considered from two aspects: On one hand, if the time resolution is too high, it would bring large statistical fluctuations, affecting the recognition of the precursor and quiescent period. On the other hand, if the resolution is too low, the structure of the quiescent period may be covered by the precursor and main emission. After all, the durations of the precursor and quiescent period are relatively short. 
     
     Next, we segment the light curve using the Bayesian Block algorithm and calculate the significance of each block using the Li-Ma formula \citep{LiMa1983}
    \begin{equation}    
    \rm{S_{LM}}=S\times{sign}(n-\alpha*b),
    \label{equ:Li-Ma_sign}
    \end{equation}

    \begin{equation} 
    \rm{S}=\sqrt{2}\left \{ n\log\left ( \frac{1+\alpha }{\alpha}\frac{n}{n+b} \right ) +b\log\left [  (\alpha +1)\frac{b}{n+b} \right ] \right \} ^{1/2},
    \label{equ:Li-Ma}
    \end{equation}

    \begin{equation}
    \rm{sign(x)}=\begin{cases}1,\quad \quad \rm{if \ x\ge 0}
    \\-1,\quad \rm{otherwise}\end{cases}
    \label{equ:sign}
    \end{equation}
    
    where $n$ is the counts of the on-source time interval $t_{on}$ to testify the existence of a burst, $b$ is the measured background counts in the off-source time interval $t_{off}$, and $\alpha$ is the ratio of the on-source time to the off-source time $t_{on}$/$t_{off}$. The first block is taken as the off-region. This is based on the assumption that the first block, which occurs before the source signal rises, represents the background region. Subsequently, we use the $\rm{S_{LM}}$ of each block to characterize the strength of the GRB signal. If none of the blocks has $\rm{S_{LM}}$ greater than 3, this GRB is considered to be too weak, and no further analysis is conducted. Conversely, for those light curves with blocks that have $\rm{S_{LM}}$ exceeding 3, all blocks are classified into low-significance blocks (LSBs) with $\rm{S_{LM}}$ below 3, and high-significance blocks (HSBs) with $\rm{S_{LM}}$ above 3. This classification analysis aims to identify candidates with proper precursor and quiescent period between the precursor and main burst.

    \subsection{Precursor and quiescent period} 
    
    First, we determine the precursor and quiescent period candidates. To identify precursor and quiescent period candidates, We find the first HSB that appears in time order and recognize it as the possible precursor; then we find consecutive LSBs preceding most HSBs and recognize these blocks as possible quiescent period. We propose three conditions to confirm quiescent period candidates:\begin{enumerate}[(1)]
    \item The quiescent period candidates should be LSBs, with the first low-significance block appearing after the first HSBs. This condition indicates that the quiescent period follows the precursor;
    \item There have to be several consecutive LSBs to determine the complete duration of quiescent period. If there are no more than one consecutive block, the first LSBs is taken;
    \item LSBs that appear after the tenth one would not be considered (which means LSBs appear at least 2s after $T_{0}$ since the light curve resolution is 0.2s, indicating a short precursor and short quiescent period.
    \end{enumerate} 
    After applying the three conditions and identifying the corresponding blocks, we sum these consecutive LSBs to determine the quiescent period candidate, while the period between the first HSB and the first LSB of the quiescent period is considered the precursor candidate.
    It is worth noting that the blocks identified above are preliminary candidates. To satisfy the standard of burst pattern of Type IL GRB, we establish a standard criterion of 2\,s for both precursor and quiescent period candidates. Specifically, if the duration of both the precursor and quiescent period candidates is less than 2\,s and the $\rm{S_{LM}}$ of the precursor candidate is not the highest among all blocks (indicating that the precursor is weaker than the main emission peak), then the precursor and quiescent period candidates are considered to satisfy the burst pattern (Fig. \ref{fig:170228}). Otherwise, these candidates are excluded.
    
    \subsection{Extended emission and dip} 
    
    After identifying the precursor and quiescent period, we establish the criterion for extended emission and dip. Extended emission typically exhibits weaker intensity and longer duration compared to the main emission. Therefore, we designate the highest $\rm{S_{LM}}$ of HSBs as representing the peak signal in the light curve and consider all HSBs that appear after the highest $\rm{S_{LM}}$ block, as well as having $\rm{S_{LM}}$ less than 1/3 of the highest $\rm{S_{LM}}$.

    In fact, the criterion of ``less than 1/3 of the highest $\rm{S_{LM}}$ block" is relatively loose. It aims to capture potential extended emission signals. Because, in the case of GRB 211211A, there are quite a few blocks whose $\rm{S_{LM}}$ is less than 1/3 of the highest $\rm{S_{LM}}$ block, although the extended emission of GRB 211211A is relatively strong. This operation yields three possible outcomes: the absence of such blocks, the discontinuous occurrence of such blocks, and at least two consecutive appearance of such blocks, which is referred to as consecutive significance blocks (CSBs).

    The first two results are disregarded. Specifically, if no such block is found, it suggests the absence of any extended emission, and if the blocks occur discontinuously, it indicates the presence of blocks with significance greater than 1/3 of the highest significance block, implying the continuation of main emission signals during those intervals. Only in the case of consecutive occurrence of such blocks (CSBs) is the extended emission candidate confirmed.

\begin{figure*}[http]
\centering
\includegraphics[width=\textwidth]{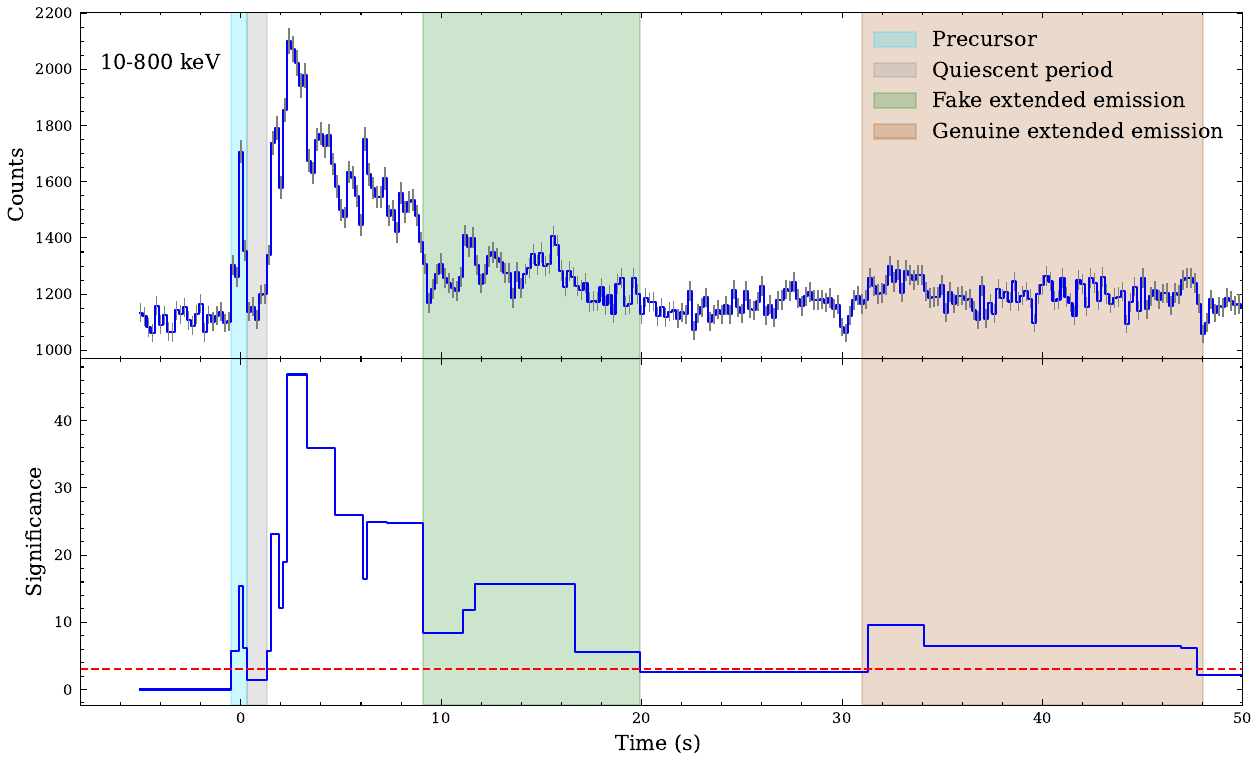}
\caption{\noindent\textbf{An example of identification of precursor, quiescent period and extended emission of GRB 170228A based on the criterion. The upper panel is the light curve of GRB, and bottom panel is the Li-Ma significance of each block partitioned by Bayesian Block. Precursor, quiescent period, fake extended emission and genuine extended emission are shaded by cyan, gray, green and tan, respectively.}}
\label{fig:170228}
\end{figure*}

    The extended emission candidate may consist of multiple segments (Fig.~\ref{fig:170228}), potentially separated by blocks of higher significance or LSBs. Similarly to the identification precursor and quiescent period, these consecutive significant blocks are initially considered as candidates for extended emission because they may appear immediately after the highest significance block, representing signals at the end of the peak rather than true extended emission (Fig.~\ref{fig:170228}). 
    
    In line with the light curve pattern of Type IL GRBs, a dip-like structure is essential to distinguish the main emission from the extended emission. Therefore, for each segment of the extended emission candidates, the blocks between the most significant block and the candidate of that segment are examined. If a block between the highest significant block and the candidate of the extended emission segment is less significant than the first block of this segment, a dip is deemed to have occurred. In this situation, candidates for extended emission are considered as genuine. Consequently, both the extended emission and the dip are found in the analysis (Fig.~ \ref{fig:170228}).

    Due to the complexity of GRB light curves, particularly the diverse profiles of extended emission, our criterion for detecting extended emission and dip is not as stringent as those for precursor and quiescent period. The process used for extended emission and dip identification aims to capture a few key blocks that clearly pertain to the extended emission. Although this criterion may not encompass all signals of extended emission, it ensures that any potential extended emission is not missed.

    In some studies, the SNR, which is defined as $S/\sqrt N$, is computed to validate the presence of extended emission, which inevitably takes into account background variations. However, in our methodology, we strive to minimize the reliance on background considerations. Apart from the initial background segment necessary for the Li-Ma significance calculation, the whole process basically does not involve background fitting or background subtraction. The reason is that the background of some GRBs in the entire dataset is quite complex, which makes uniform background subtraction challenging and potentially introduces uncertainties.

    \begin{table*}[htbp]
    \caption{\centering{Timescale of different phases of 5 GRBs}}
    \begin{tabular*}{\hsize}{@{}@{\extracolsep{\fill}}ccccc@{}}
    % \hline
    \toprule
    GRB name& $T_{90}$(s) & $T_{\rm pre}$(s) & $T_{wt}$(s) & $T_{EE}$(s)\\
    \hline
    GRB 230307A & 41.52 & 0.40 & 0.30 & 50.80\\
    GRB 211211A & 43.18 & 0.20 & 0.93 & 50.80\\
    GRB 200914A & 65.28 & 0.40 & 1.20 & 49.50\\
    GRB 200311A & 52.48 & 0.40 & 0.20 & 39.50\\
    GRB 170228A & 60.20 & 0.80 & 1.20 & 44.20\\
    \botrule
    \end{tabular*}
    \label{tab:timescale_of_GRBs}
    \end{table*}
    
\section{Search Results and Verification}\label{section4}
    
\begin{figure*}[http]
\centering
\includegraphics[width=\textwidth]{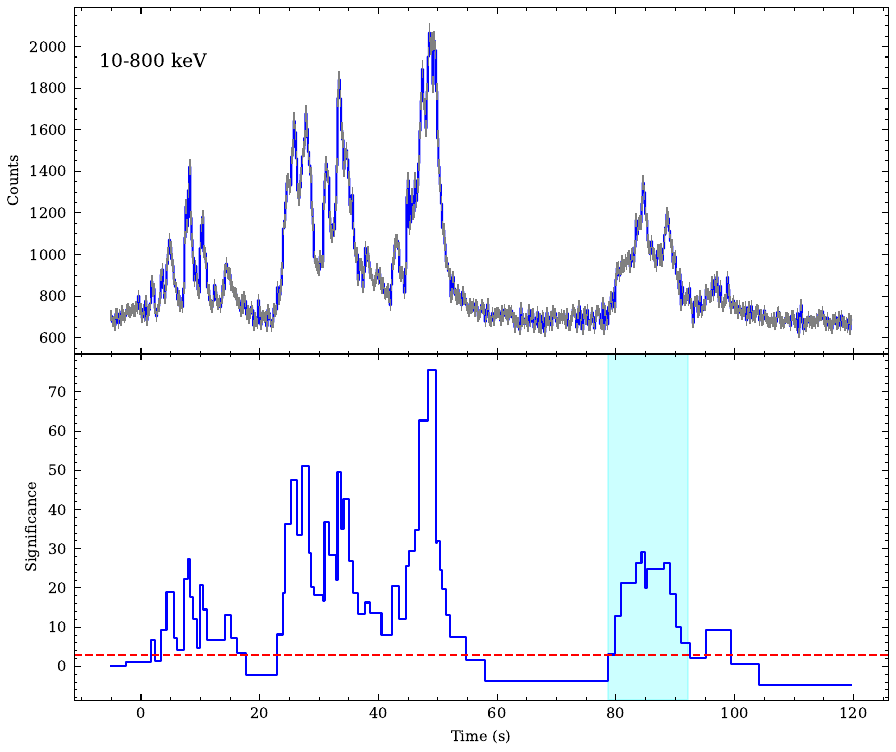}
\caption{\noindent\textbf{Identification of extended emission of GRB 140810A based on the criterion. The upper panel is the light curve of GRB, and bottom panel is the Li-Ma significance of each block partitioned by Bayesian Block. The extended emission are shaded by cyan.}}
\label{fig:140810}
\end{figure*}

\begin{figure*}[http]
\centering
\includegraphics[width=\textwidth]{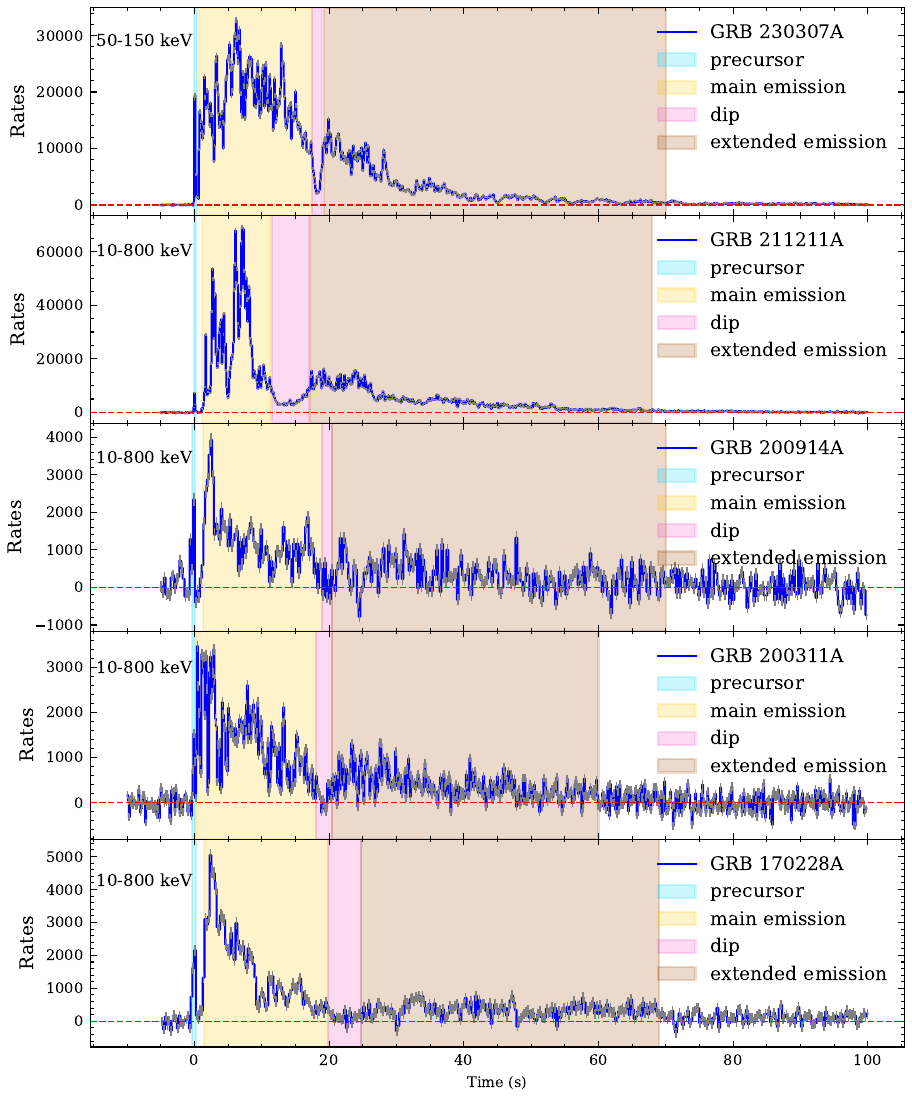}
\caption{\noindent\textbf{The background subtracted lightcurves of GRB 230307A,  GRB 211211A, GRB 170228A, GRB 200914A and GRB 200311A.  }The light curve of GRB 230307A is obtained from GECAM with GRD01, GRD04, GRD05 summed, the light curve of GRB 211211A is obtained from GBM with n2 and na summed,  the light curve of GRB 170228A is obtained from GBM with n0,n1,n2,n9,na summed, the light curve of GRB 200914A is obtained from GBM with n0,n1,n3,n4,n5 summed, the light curve of GRB 200311A is obtained from GBM with n4,n7,n8 summed. The shaded intervals represent precursor, main emission, dip and extended emission of each light curve structure.}
\label{fig:five_lightcurves}
\end{figure*}

    Applying the criterion mentioned above, we found 20 GRBs satisfying the precursor and quiescent period criterion among the 1819 GRBs in the past decade, and five of them satisfying the extended emission and dip criterion. Among the five GRBs, for the GRB 140810A, it is worth noting that the dip duration of this burst is very long, much longer than that of the other Type IL GRBs, suggesting that the central engine of this burst may have stopped prompt activity during the dip stage and entered the afterglow stage. Therefore, the emission after dip is more likely to be a flare rather than part of prompt emission, whereas the criteria proposed in this paper aim to prompt emission.
    % the signal after the main emission looks more like a flare than a persistent extended emission, 
    So we exclude it (Fig.~\ref{fig:140810}). The reason why we do not include dip duration as one of the criterion is that dip has not been systematically studied like precursor and quiescent period, as well as some correlations. Therefore, according to the criterion, we get four GRBs that satisfy the light curve pattern. In addition, GRB 230307A does not satisfy the criterion, mainly because it is so bright and the signal in ``quiescent period" is obviously higher than the background. However, just as analyzed in \cite{wangtype1L}, after weakening GRB 230307A to the strength of GRB 211211A, it also shows a real quiescent period and satisfy the criterion, thereby proving that GRB 230307A also belongs to Type IL GRB. So we end up with a sample of Type IL GRB with five bursts, i.e. GRB 230307A, GRB 211211A, GRB 200914A, GRB 200311A and GRB 170228A (Fig.~\ref{fig:five_lightcurves}).

    % which has been proved in the \cite{wangtype1L} that by reducing its brightness and examining its precursor, GRB 230307A is still a Type IL GRB. 
    
    \subsection{Basic information of these GRBs} 

     GRB 211211A was detected at 2021-12-11T13:09:59.651 (UTC) by \textit{Fermi}/GBM, Swift/BAT, \textit{Insight}-HXMT and other telescopes \citep{11AGBMGCN,11AbatGCN,11AhxmtGCN}. The precursor of this GRB lasts about 0.2\,s with a surprising QPO feature \citep{XS_GRB211211A_QPO}. The emission after the precursor is clearly divided into two parts, a long and hard main emission and a soft and long extended emission \citep[see e.g.,][]{YJ_GRB211211A_nature,11A_MVT,XS_GRB211211A_QPO}. The $T_{90}$ of GRB 211211A is up to about 50\,s.
    
    GRB 230307A is the second brightest GRB ever recorded \citep{2025NSRSun}. On 2023 March 7, 15:44:06.65 UT ($T_0$), the $GECAM-B$ was triggered in-flight by this exceptionally bright long burst, which is also detected by many other missions such as \textit{Fermi}/GBM \citep{Fermi2023GCN} and Konus-WIND \citep{Konus2023GCN}. The extreme brightness of this burst was first reported to the community by GECAM-B with the real-time alert data \citep{xiong2023GCN}, and subsequently confirmed by other instruments, leading to a large observation campaign for this event. The first pulse of GRB 230307A was proved to be the precursor \citep{wangtype1L}, and the main emission and the extended emission are quite clear from the observation of the light curve. The $T_{90}$ of GRB 230307A is up to about 50\,s. The light curve extraction for GECAM was performed using the GECAMTools-v20240514\footnote{\url{https://github.com/zhangpeng-sci/GECAMTools-Public}}.
    
    GRB 200914A was detected on 2020-09-14T12:48:30 (UTC) by \textit{Fermi}/GBM \citep{0914GBMGCN}, and the observations of this burst are also reported by Konus-WIND \citep{0914KWGCN} and INTEGRAL \citep{0914INTEGRALGCN}. The $T_{90}$ of GRB 200914A is up to about 65\,s. 
    
    GRB 200311A was detected on 2020-03-11T15:16:12 (UTC) by \textit{Fermi}/GBM \citep{0311GBMGCN}, and the observations of this burst are also reported by Konus-WIND \citep{0311KWGCN}. The $T_{90}$ of GRB 200311A is up to about 52\,s. 

    GRB 170228A was detected on 2017-02-28T19:03:00.17 (UTC) by \textit{Fermi}/GBM \citep{0228GBMGCN}, and the observations of this burst are also reported by Konus-WIND \citep{0228KWGCN} and \textit{Fermi}/LAT \citep{0228LATGCN}. The $T_{90}$ of GRB 170228A is up to about 60\,s.
    
    Similar to the first two GRBs (GRB 230307A and GRB 211211A), the latter three GRBs can also be segmented into three distinct episodes: the precursor, which is relatively significant over the background; the quiescent period subsequent to the precursor; and a bright main emission, followed by a persistent extended emission with a dip separating it from main emission. The durations of each episode of these five GRBs are listed in Table~\ref{tab:timescale_of_GRBs}.

    We find that the light curves of these GRBs exhibit a multi-peak structure, and the occurrences of these GRBs are not evenly distributed over time.     
    Unfortunately, there is no publicly available follow-up x-ray or optical observation for the latter three GRBs, hence it is unknown whether there are any kilonova signals, and the redshifts of them are not measured. 
    
    \subsection{Verification of Type IL GRB} 
    
    Since no x-ray afterglow has been detected in the latter three GRBs, we can only demonstrate their nature of Type IL GRB by the observation properties of prompt emission.
    
    \subsubsection{Belonging to Type I GRB}

    Traditionally, the Amati relation mostly contains Type II GRBs, while Type I GRBs were regarded as outliers of the correlation. Type I GRBs are found to form a class distinct from Type II GRBs. Therefore, the Amati relation is widely used to classify Type I GRB and Type II GRB because of its significant distinction between the two types of GRBs. Based on the correlation, some studies integrated E$_{p,i}$ and E$_{iso}$ from the Amati relation into a single parameter, considering the duration, thus proposing a new classification method \citep{MinaevHR2020,HouJun13}. 
    
\begin{figure*}
\centering
\begin{tabular}{cc}
\begin{overpic}[width=0.45\textwidth]{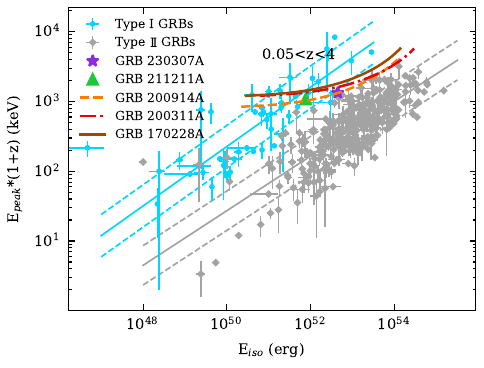}\put(0, 70){\bf a}\end{overpic} &
        \begin{overpic}[width=0.45\textwidth]{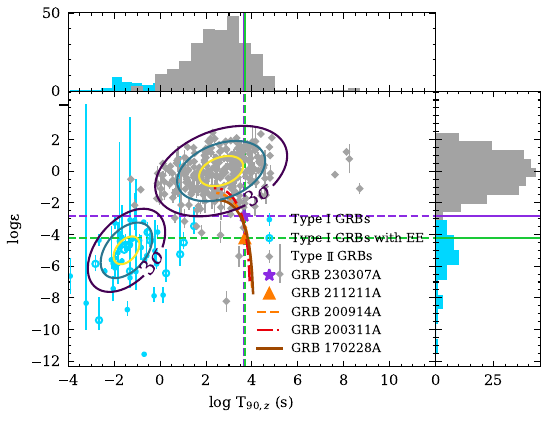}\put(0, 70){\bf b}\end{overpic} \\
\begin{overpic}[width=0.45\textwidth]{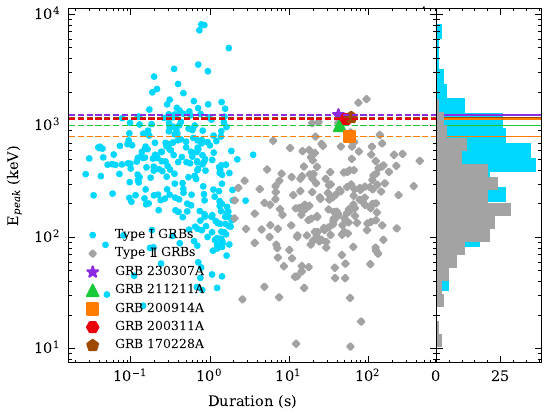}\put(0, 70){\bf c}\end{overpic} &
        \begin{overpic}[width=0.45\textwidth]{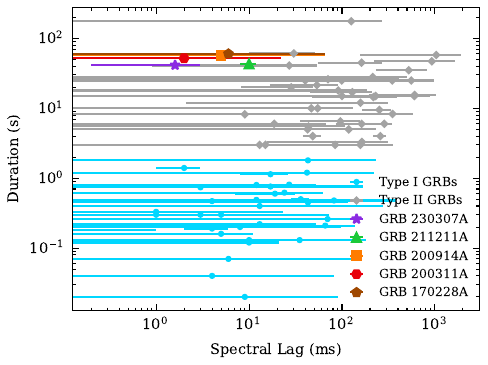}\put(0, 70){\bf d}\end{overpic} \\
\begin{overpic}[width=0.45\textwidth]{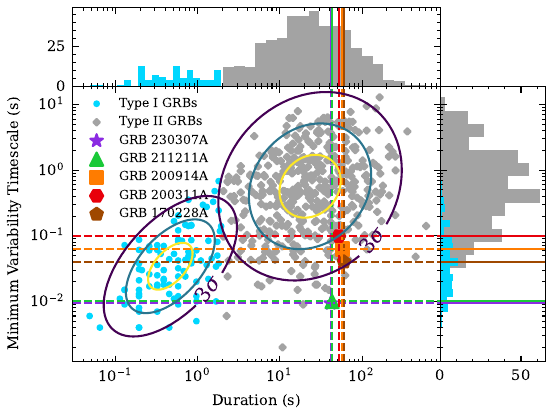}\put(0, 70){\bf e}\end{overpic}&
        \begin{overpic}[width=0.45\textwidth]{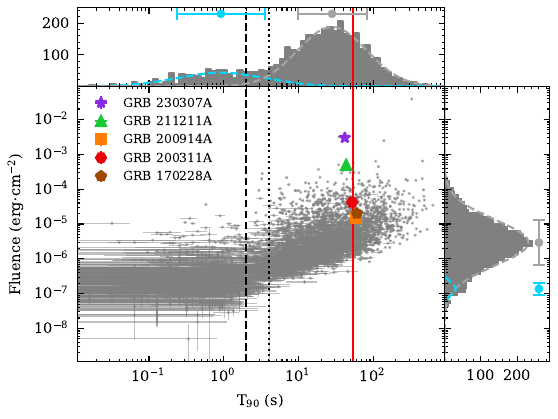}\put(0, 70){\bf f}\end{overpic} \\

\end{tabular}
\caption{\noindent\textbf{Position of Type IL GRB samples in various GRB classification schemes.} \textbf{a}, The E$_{p,z}$ and E$_{\rm iso}$ correlation diagram. The best-fit for Type II and Type I GRBs are plotted (solid lines) with the 1$\sigma$ boundary (dashed line) marked.  The GRB sample is from the literature \citep{Lan23}. \textbf{b}, The $log\varepsilon$ versus intrinsic duration diagram. The possibility contours for GRB clustering are marked with 1$\sigma$ (yellow), 2$\sigma$ (blue) and 3$\sigma$ (purple) respectively. The GRB sample is from the literature \citep{MinaevHR2020}. In \textbf{a-b}, The yellow dashed line, red dashdot line and brown solid line represent the evolution of the E$_{p,z}$ and E$_{\rm iso}$ of GRB 200914A, GRB 200311A and GRB 170228A as the redshift changes from 0.05 to 4, respectively. \textbf{c}, The peak energy versus duration diagram. The GRB sample is from the literature \citep{2017LusGRBEp,2021GBMcat,2017KWcat}.  \textbf{d}, The duration versus lag diagram. The GRB sample is from the literature \citep{Bernardinilag2015,GehrelsLAG2006,Goldstein0817,Xiaolag2022}. \textbf{e}, The minimum variability timescale versus duration diagram. The possibility contours for GRB clustering are marked with 1$\sigma$ (yellow), 2$\sigma$ (blue) and 3$\sigma$ (purple) respectively. The GRB sample is from the literature \citep{GolkhouMVT2015}. \textbf{f}, The fluence versus duration diagram. The best-fit using double-gaussian function for the fluence and duration histgram and their 1$\sigma$ errorbars are plotted. The dashed line, dotted line and solid line represent 2\,s, 1$\sigma$ of the duration distribution (4\,s) and  3$\sigma$ of the duration distribution (52.8\,s), respectively. All error bars on data points represent their 1$\sigma$ confidence level.}
\label{fig:classification}
\end{figure*}

    Here we adopt the parameter $\varepsilon = E_{\gamma ,iso,52}/E_{p,z,2}^{5/3}$ from \cite{HouJun13}, and plot the Amati relation and $log\varepsilon$ - $logT_{90,z} $ diagram (Fig.~\ref{fig:classification}a,b). It shows that both GRBs with determined redshift, GRB 230307A and GRB 211211A, fall in Type I GRB region, while the three GRBs without redshift measurements, GRB 200914A, GRB 200311A and GRB 170228A fall largely in Type I GRB region as the redshift varies. This clearly indicates that the five Type IL GRBs all belong to Type I GRBs in terms of spectral and energetic characteristics.

    Since Type I GRBs are more likely to occur in high magnetic environments, the $E_{peak}$ of Type I GRBs is relatively higher than that of Type II GRBs. We measured $E_{peak}$ of five Type IL GRBs using Pyxspec software \citep{pyxspec} and found that their $E_{peak}$ is closer to that of Type I GRBs compared to Type II GRBs (see Fig.~\ref{fig:classification}c) .

    Spectral lag, a parameter that characterizes temporal-spectral property, was also used to classify GRBs \citep{GehrelsLAG2006}. Type I GRBs normally have negligible spectral lags, whereas larger positive spectral lags are characteristic of Type II GRBs. We calculate the spectral lag for these Type IL GRBs following treatments in \cite{Ukwattalag2012} and find that the spectral lag of these Type IL GRBs are consistent with Type I GRBs (see Fig.~\ref{fig:classification}d).

    In addition to their spectral properties, the time variability properties can also help distinguish Type I GRBs from Type II GRBs. Since the emission region of Type I GRBs is smaller than that of Type II GRBs, Type I GRBs exhibit a smaller minimum variability timescale (MVT) \citep{GolkhouMVT2015}. We calculated the MVT of Type IL GRBs using Bayesian blocks and found that they also fall within a region closer to that of Type I GRBs rather than Type II GRBs (see Fig.~\ref{fig:classification}e).

    In summary, we confirmed that the four GRBs identified using the criterion, as well as GRB 230307A, all belong to Type I GRBs based on their spectral and time variability properties of the prompt emission.

    \subsubsection{Forming a distinct subclass}

    After confirming that these five GRBs belong to Type I GRBs, we now demonstrate that they form a distinct subclass. 

    First, we look back at the position of their durations in the GRB sample. We obtain all GRBs from the Fermi GBM Burst Catalog and fit their durations and fluence with a double-log-Gaussian function (Fig.~\ref{fig:classification}f). If the distribution of long and short bursts is considered to be continuous rather than truncated at 2\,s, then the durations of the five Type IL GRBs are basically on the $3\sigma$ edge of the short burst distribution. According to the short burst sample, the expected number of Type I GRBs with a duration greater than $3\sigma$ of the distribution is about 0.8 in the last decade, that is, the number of Type IL GRBs we found is significantly larger than the expected number. This fact indicates that Type IL GRB form a distinct class from Type I GRBs.

\begin{figure*}
\centering
\begin{tabular}{c}
\begin{overpic}[width=0.5\textwidth]{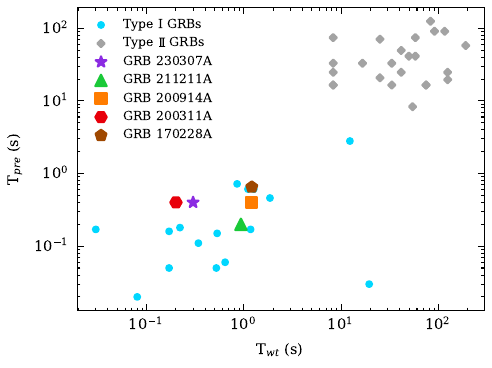}\put(0, 70){\bf a}\end{overpic} \\
        \begin{overpic}[width=0.5\textwidth]{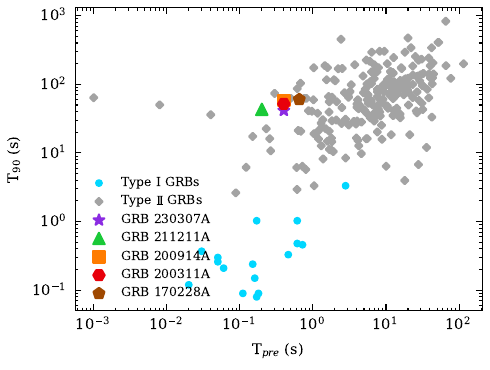}\put(0, 70){\bf b}\end{overpic} \\
        \begin{overpic}[width=0.5\textwidth]{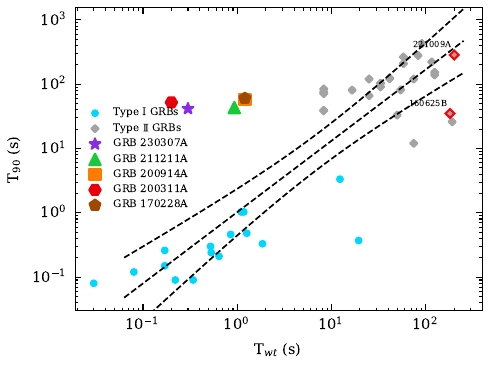}\put(0, 70){\bf c}\end{overpic} \\
\end{tabular}
\caption{\noindent\textbf{The temporal properties of Type IL GRB samples.} \textbf{a}, The $T_{90}$ vs T$_{pre}$ diagram. \textbf{b}, The T$_{pre}$ vs $T_{wt}$ diagram. \textbf{c}, The $T_{90}$ vs $T_{wt}$ diagram. The dashed black line represents the power-law fitting and the 3$\sigma$ error of the $T_{90}$ and $T_{wt}$.} In \textbf{a}--\textbf{c}, Type I and Type II GRBs are represented by cyan solid circles and gray solid diamonds, respectively. In \textbf{c}, two typical Type II GRBs with extended emission, GRB 221009A and GRB 160625B are put in red diamonds. The GRB sample is obtained from the literature \citep{Wangpre2020,Hupre2014}.
\label{fig:duration}
\end{figure*}

    Then we demonstrate that they form a distinct subclass subclass from the perspective of light curve characteristics. We found that in samples summarized in previous studies, the precursors and quiescent periods for Type II GRBs are generally longer, while those for Type I GRBs are relatively short \citep{Wangpre2020}. From the perspective of the light curve pattern, Type IL GRBs have moderate precursor and short quiescent period, making them are closer to Type I GRB (short burst) in the samples of Fig.~\ref{fig:duration}a, and at the boundary of Type I GRB (short burst) and Type II GRB (long burst) in the samples of Fig.~\ref{fig:duration}b. However, for these five Type IL GRBs, they all locate in the same region to form a separate cluster of GRB with a short quiescent period but long duration, making themselves a distinct class in $T_{90}$-$T_{wt}$ distribution map (Fig.~\ref{fig:duration}c). For the samples in Fig.~\ref{fig:duration}c, the pearman coefficient between $T_{90}$-$T_{wt}$ in log space is 0.85 with the $p$-value being 3.7e-12, so we perform a power-law fit to the relation of the $T_{90}$-$T_{wt}$ of samples and plot the 3 $\sigma$ error of the fitting. It shows that five Type IL GRBs fall far from the 3 $\sigma$ region and group together, which further proves that the five Type IL GRBs belong to the distinct class.

    In addition, apart from expected number and light curve characteristics, some other properties of Type IL GRBs illustrate that outside the framework of the criterion, they also form a distinct subclass. First, the distribution of the duration of main emission in these 5 GRBs is relatively concentrated. Secondly, the duration of these five Type IL GRBs mainly contributed by main emission rather than extended emission, resulting in the energy ratio between extended emission and main emission is almost constant (see \ref{section443}). These properties distinguish Type IL GRB from the traditional ``long-short” GRB, which is not required in the criterion, therefore, making Type IL GRB a distinct subclass.
    
    \subsubsection{Different from typical ``long-short" GRB}\label{section443}

    Since GRB 060614, GRBs with similar characteristics and properties (including GRB 211227A) have been referred to as ``long" short bursts. However, we would like to clarify that Type IL GRB discussed in our paper is not the same as these traditional ``long" short bursts.

\begin{figure*}
\centering
\begin{tabular}{cc}
\begin{overpic}[width=0.45\textwidth]{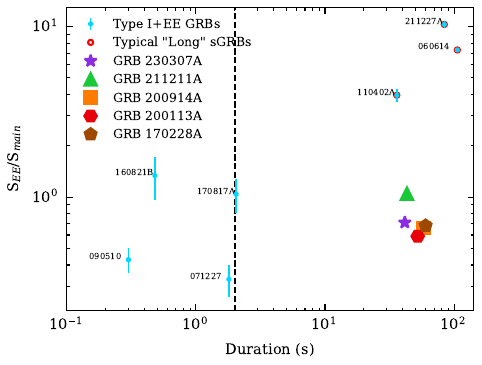}\put(0, 70){\bf a}\end{overpic} &
        \begin{overpic}[width=0.45\textwidth]{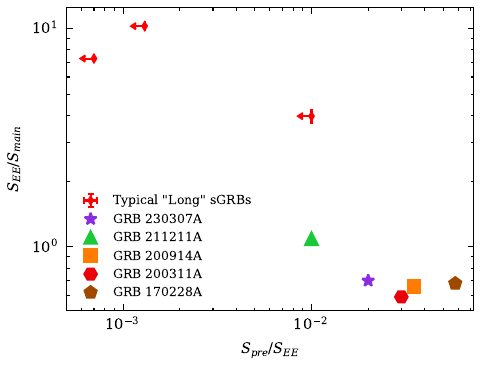}\put(0, 70){\bf b}\end{overpic}

\end{tabular}
\caption{\noindent\textbf{Comparison of Type IL GRB with typical ``long" sGRB.} 
\textbf{a}, The ratio of photon fluence from 15\,keV to 100\,keV between the extended emission and the main emission ($S_{EE}/S_{main}$) versus duration diagram. The red hollow circles represent the typicle ``long'' short GRBs, GRB 060614, GRB 211227A and GRB 110402A. 
\textbf{b}, ($S_{EE}/S_{main}$) versus the ratio of photon fluence between precursor and extended emission $S_{pre}/S_{EE}$. The leftward arrows represent the 3$\sigma$ upper limits of $S_{pre}/S_{EE}$ of typical ``long" sGRB without obvious precursors.}
\label{fig:ratio}
\end{figure*}

    Except for the existence of precursor, the difference between Type IL GRB and ``long" short bursts mainly lies in the main emission, rather than extended emission. We study the ratio of photon fluence between extended emission and main emission ($S_{\rm EE}/S_{\rm main}$). As depicted in Fig.~\ref{fig:ratio}a, Type IL GRBs have $S_{\rm EE}/S_{\rm main}$ close to unity and a relatively concentrated distribution, which means that the photon fluence of the extended emission is comparable to that of the main emission, while $S_{\rm EE}/S_{\rm main}$ of these ``long" short GRBs are relatively large and have a scattered distribution. This indicates that the long duration of Type IL GRBs is not primarily contributed by the soft long extended emission but by the hard long main emission, which is the most significant difference from the case in ``long" short bursts. Therefore, the ratio of $S_{\rm EE}/S_{\rm main}$ significantly distinguishes Type IL GRBs from classical ``long" short GRBs. 
       
    Additionally, we further investigate the photon fluence ratio between of the precursor ($S_{\rm pre}$) and extended emission for Type IL GRBs and ``long" short GRBs respectively. Since there is no precursor detection in typical ``long" short GRB, we calculate the 3 sigma upper limits of the precursor fluence and then determine the $S_{\rm pre}/S_{\rm EE}$ for them. As depicted in Fig.~\ref{fig:ratio}b, $S_{\rm pre}/S_{\rm EE}$ of Type IL GRB is much larger that of typical ``long" short GRB, indicating that the relative brightness of precusor is another difference between Type IL GRB and typical ``long" short GRB.

    \begin{table*}[htbp]
    \caption{\centering{The spectral analysis results of three GRBs}}
    \begin{tabular*}{\hsize}{@{}@{\extracolsep{\fill}}ccc|ccc|cc|cccc@{}}
    % \hline
    \toprule
    GRB & Phase & t$_{1}$$\sim$t$_{2}(s)$ && {CPL} &&{BB} &&& {CPL+BB} \\
    &&&$\alpha$ & $E_{cut}$(keV) & BIC & kT(keV) &BIC & $\alpha$ & $E_{cut}$(keV) & kT(keV)  & BIC\\
    \hline
    200914A & pre & -0.3 $\sim$ 0.1& -0.74$^{+0.42}_{-0.26}$ & 835$^{+1630}_{-549}$ & 115.07 & 69.9$^{+11.2}_{-9.8}$ & 113.75 &  &  Not-Con. &  & \\
    &ME &  0.3 $\sim$ 19 & -0.78$^{+0.04}_{-0.03}$ & 930$^{+120}_{-106}$ & 239.17 & &&&&& \\
    &Tot &  -0.3 $\sim$ 65 & -0.94$^{+0.03}_{-0.04}$ & 906$^{+167}_{-124}$ & 403.97 & &&&&& \\
    200311A & pre & -0.3 $\sim$ 0.1& -1.46$^{+0.14}_{-0.10}$ & 2340$^{+1660}_{-1470}$ & 81.15 & 25.50$^{+3.76}_{-3.07}$ & 89.24 & &  Not-Con. &  & \\
    &ME &  0.3 $\sim$ 18 & -0.66$^{+0.03}_{-0.03}$ & 856$^{+69.7}_{-65.7}$ & 124.63 & &&&&& \\
    &Tot &  -0.3 $\sim$ 52 & -0.88$^{+0.02}_{-0.03}$ & 1020$^{+102}_{-92.2}$ & 205.47 & &&&&& \\
    170228A & pre & -0.5 $\sim$ 0.3& -1.05$^{+0.20}_{-0.32}$ & 1160$^{+1770}_{-643}$ & 74.61 & 57.10$^{+7.75}_{-5.94}$ & 89.88 & -1.12$^{+0.13}_{-0.10}$ & 2160 $^{+1490}_{-1210}$ & 30.3$^{+18.2}_{-23.5}$ & 82.84\\
    &ME &  1.4 $\sim$ 20 & -0.64$^{+0.05}_{-0.04}$ & 606$^{+71.1}_{-59.3}$ & 101.07 & &&&&& \\
    &Tot &  -0.5 $\sim$ 60 & -1.00$^{+0.04}_{-0.04}$ & 1040$^{+197}_{-164}$ & 182.78 & &&&&& \\
    
    \botrule
    \end{tabular*}
    \label{tab:spectral_of_GRBs}
    \end{table*} 
    
    \section{Discussions}\label{section5}

    In this section, we discuss the physical implications of Type IL GRB. First, we search for the thermal component in the precursor of Type IL GRBs since there are reported thermal component in the precursor of some Type I GRBs \citep{Wangpre2020}. Furthermore, based on properties of the main emission and extended emission of Type IL GRBs, we discuss several prevalent theoretical models for progenitor, central engine and extended emission of Type IL GRBs. Our intention is not to definitively confirm or exclude any specific model, as such an endeavor is beyond the scope of this paper; rather, we are merely discussing the models to which Type IL GRBs are more favored.

    The first property of Type IL GRB is the long duration of the main emission, which generally lasts about 10\,s. Therefore, it is necessary to consider what is the progenitor of this Type of GRB and how the central engine produced by the merger can remain active for such a long time, along with the related question: How should the energy of the main emission and extended emission be distributed?
    
    The second property is that, for typical ``long" short bursts, the range of variation in the ratio of photon fluence between the main burst and extended emission is large \citep{NorrisshortEE2006}, indicating that the components are physically decoupled or that the physical model must explain the reasons for this wide range. In contrast, this ratio for Type IL GRBs is almost constant ($\sim0.7$, with small scattering), suggesting that these two components are closely related to each other,
    which has a strong implication on the mechanism of Type IL burst.
    
    The third property is that most extended emissions of merger bursts (short GRB) appear platform-like or hump-shaped, while the extended emission of several Type IL GRB in this paper feature both platform-like structures and power-law decay, thus can be fitted by the  smoothed broken power-law (SBPL) model (Fig.~\ref{fig:EEflux&HLE}a). Hence, for those models which can produce power-law decaying extended emission would be more preferred for Type IL GRB. 

    \subsection{Precursor}

    Since the thermal and non-thermal component of GRB 230307A and GRB 211211A have been determined in\cite{wangtype1L,XS_GRB211211A_QPO} respectively, we focus on the remaining three Type IL GRBs found in this work. We use the cutoff power-law (CPL) model, blackbody (BB) model and CPL+BB model to fit the precursor spectrum, and use the Bayesian information criterion (BIC) to evaluate the goodness of fit of these models, where 2 $\le$ $\Delta$BIC  \textless  6 gives positive evidence, and $\Delta$BIC $\ge$ 6 gives strong positive evidence in favor of the model with a lower BIC \citep{BIC}. 
    
\begin{figure*}
\centering
\begin{tabular}{cc}
\begin{overpic}[width=0.45\textwidth]{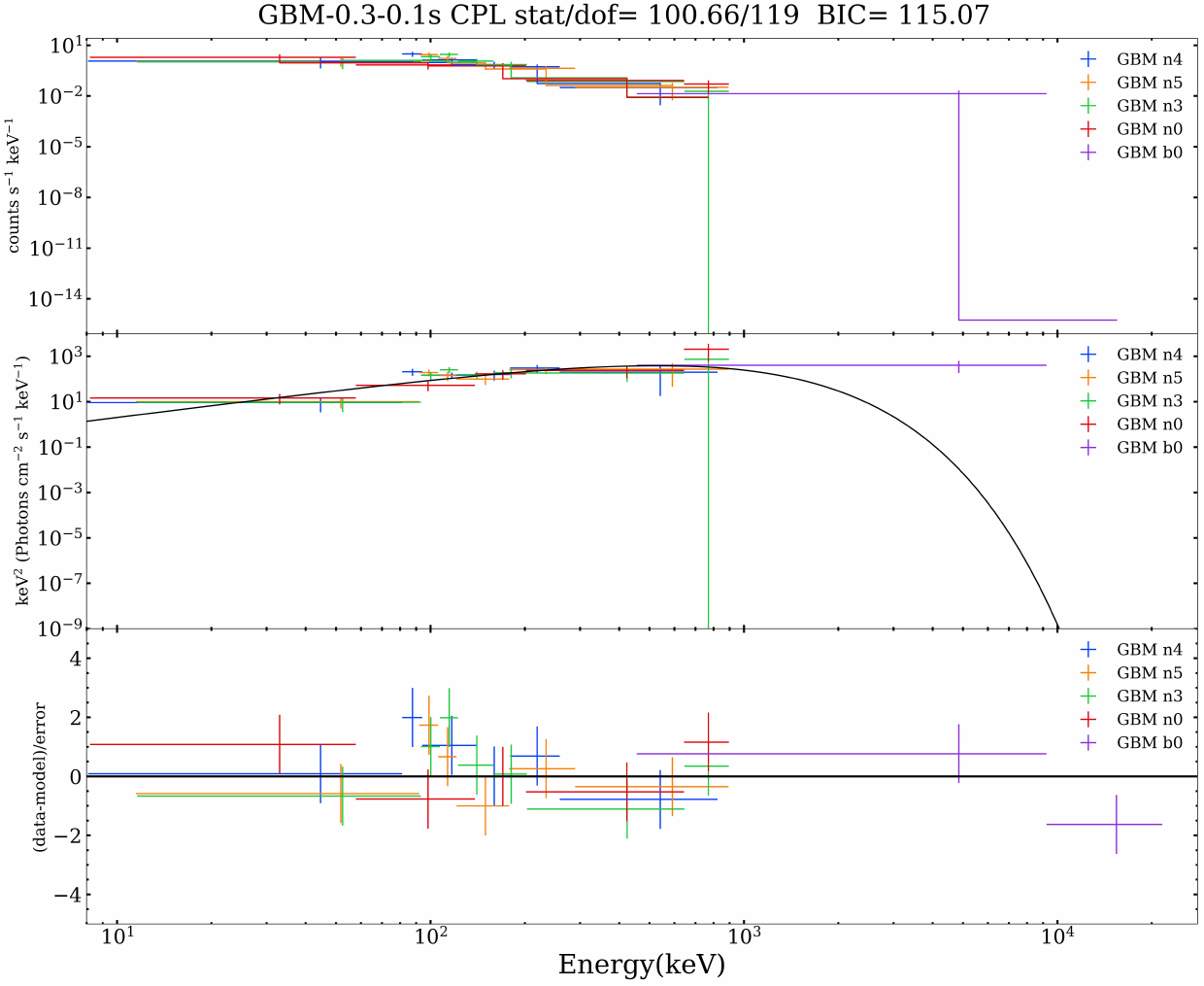}\put(0, 80){\bf a}\end{overpic} &
        \begin{overpic}[width=0.45\textwidth]{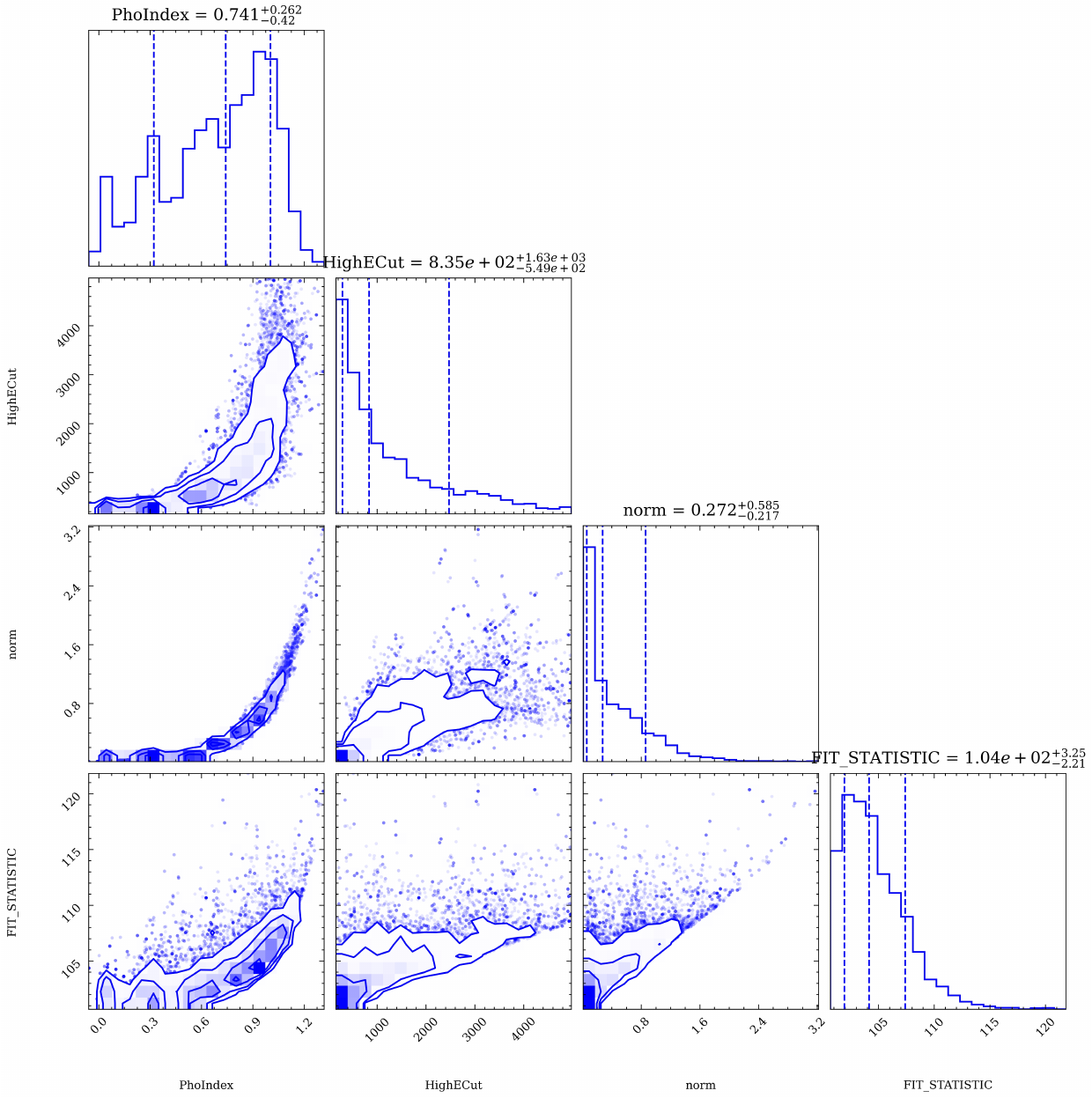}\put(0, 80){\bf b}\end{overpic} \\
\begin{overpic}[width=0.45\textwidth]{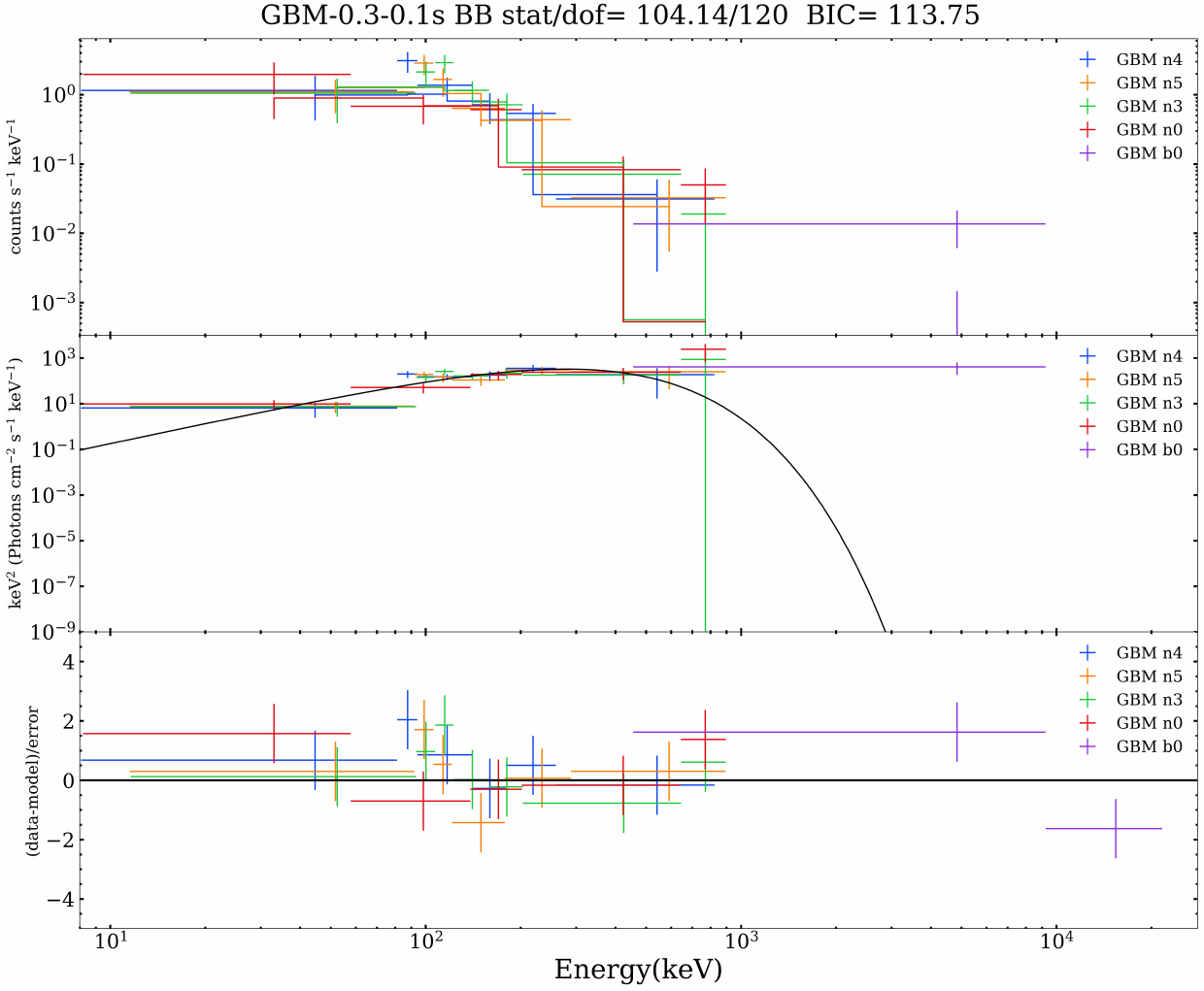}\put(0, 80){\bf c}\end{overpic} &
        \begin{overpic}[width=0.45\textwidth]{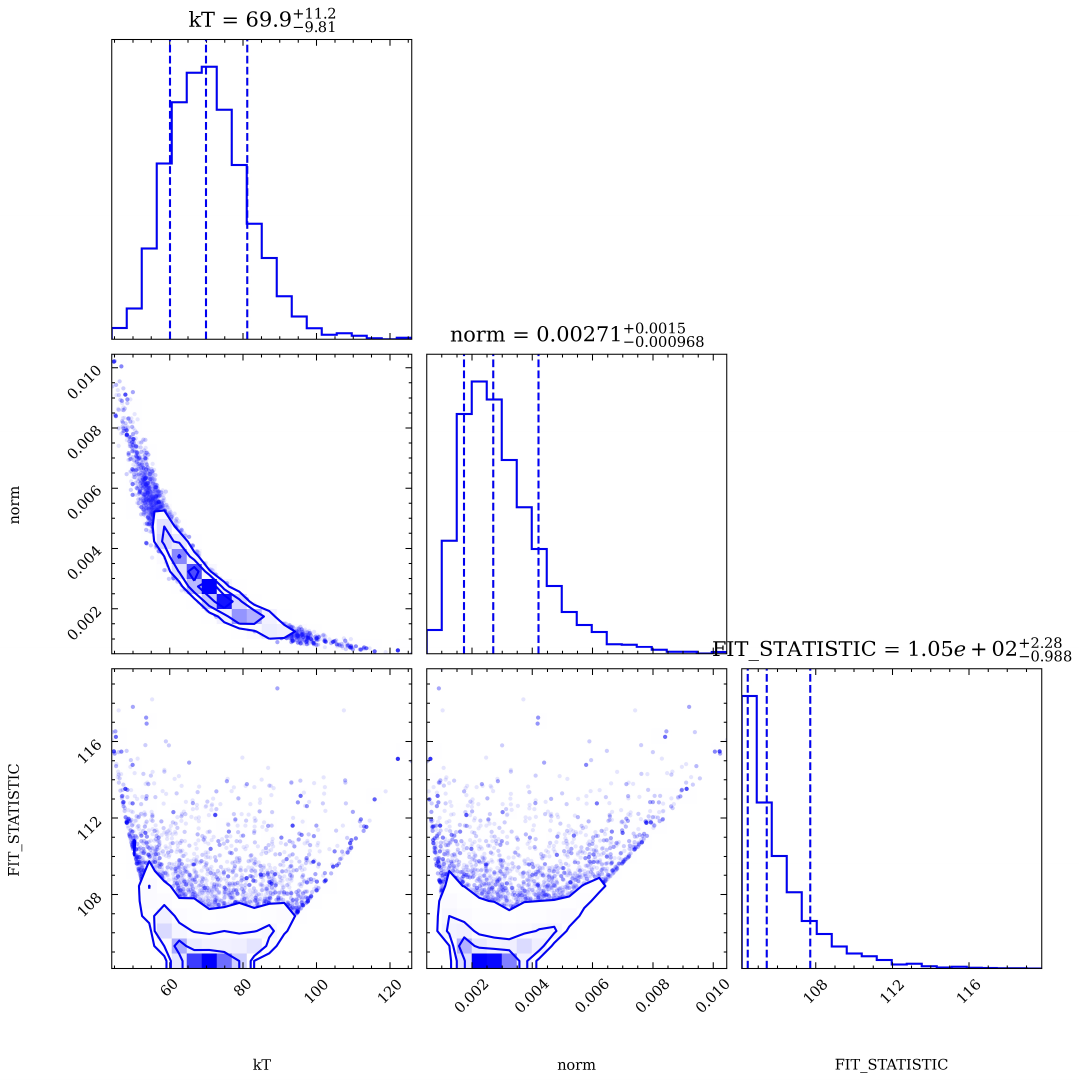}\put(0, 80){\bf d}\end{overpic} \\
\end{tabular}
\caption{\noindent\textbf{Spectra fitting results of the precursor of GRB 200914A.} 
\textbf{a-b}, the CPL model (left panel) and corner plot of the posterior probability distributions of the parameters (right panel).  
\textbf{c-d}, the BB model (left panel) and corner plot of the posterior probability distributions of the parameters (right panel). }
\label{fig:0914_pre_fitting}
\end{figure*}

    We find that for the GRB 170228A and GRB 200311A, the precursor is relatively hard and, therefore, exhibits a broad spectrum. Therefore, the CPL model can be well constrained and has the lowest BIC. However, for GRB 200914A, the precursor has a soft and narrow spectrum, hence the BB model has slightly lower BIC than that of CPL. Combined with the photon index, which is close to the death line of synchrotron radiation, it is reasonable to consider that there is a thermal component in the precursor of GRB 200914A (Fig.~\ref{fig:0914_pre_fitting}) (see Table~\ref{tab:spectral_of_GRBs}). It should be noted that the thermal component does not necessarily have to be present in the observational data: as discussed in \citep{wangtype1L}, when the brightness is sufficiently low, the thermal component may not be significant in the energy spectrum.

    \subsection{Progenitor and central engine}
    
    The central engine of GRB 230307A and GRB 211211A is proposed to be magnetar \citep{2025NSRSun,YJ_GRB211211A_nature}. The extended emission in the gamma-ray energy range of GRB 230307A can be explained by high-latitude effects (HLE), while the flux in the X-ray energy range is explained by the spin-down radiation of the magnetar \citep{2025NSRSun}. GRB 170228A, GRB 200914A and GRB 200311A do not comply with the high-latitude effects in the gamma-ray energy range (Fig.~\ref{fig:EEflux&HLE}b), thus requiring additional components to explain the extended emission. Considering the prolonged emission, and the fact that the occurrence rate of Type IL GRB within Type I GRB (about 5/400) is comparable to the proportion of magnetars among neutron stars ($\sim1\%$) \citep{Kaspi2017Magnetars}, it is reasonable to consider the magnetar as the central engine.

    If the accretion-powered engine is considered, the duration of main emission could by defined as\citep{2025zhang}:
    
    \begin{equation}    
    T_{GRB}\simeq  max(t_{ff},t_{acc})-t_{bo} ,
    \label{equ:tgrb}
    \end{equation}
    
    where $t_{ff}$ is free-fall timescale of the progenitor star, $t_{\rm ff}\sim\left( \frac{3\pi}{32G\rho} \right)^{1/2}$, where $\rho$ is the density of progenitor star \citep{2018GRBbook}. The $t_{acc}$ is the  characteristic timescale for accretion, which is the viscous timescale $t_{visc} = R^{2}/\alpha \Omega _{k}H^{2}$, given by 

    \begin{equation}    
    t_{visc} \approx 0.2s\ \left ( \frac{M}{M_{\odot } } \right )^{-1/2}\ \left ( \frac{0.1}{\alpha } \right )\ \left(\frac{R_{0} }{4R_{NS} }\right )^{3/2}\ \left(\frac{H/R_{0}}{0.2}\right)^{-2},
    \label{equ:tvisc}
    \end{equation}
    
    where $R_{NS}\sim 10\ km$ is the typical value of NS radius, $\alpha$ is the viscosity parameter and $M$ is the mass of magnetar; $R_{0}$, $H$ and $\Omega _{k} \equiv ({GM}/{R_{0}^{3} } )^{\frac{1}{2} }$ are the disc’s radius, disk thickness and Keplerian rotation rate, respectively. The ${H/R_{0}}$ is taken as 0.2 within the neutrino cooled disc accreting at $\dot{M}\simeq 0.1\sim 1M_{\odot } s^{-1} $ \citep{Chenacc2007}. The $t_{bo}$ is the timescale for the jet to break out the envelope of the progenitor, which is usually taken as $\sim 1s $ for a compact binary merger remnant. 
    
    In this scenario, as one can see, the characteristic timescale for accretion is relatively small, so we need to consider the duration of main emission from the perspective of progenitor. The involvement of progenitor system with a white dwarf (WD) seems to be reasonable, as the density of a WD is smaller than that of a NS, resulting in a longer free-fall timescale, which is comparable to the long duration of the main emission. This is similar to the explanation for GRB 211211A \citep{YJ_GRB211211A_nature}.

    In the binary merger system involving WD, \cite{Chenjunping24} proposed that during the coalescence process, the repeated partial disruptions (RPDs) of the WD would take place by the NS (or the BH). This process would likely modulate the luminosity variation of an accretion driven jet, resulting in light curves exhibiting weak periodicity, known as quasi periodic modulation (QPM). Moreover, GRB 230307A and GRB 211211A did exhibit relatively weak periodic activity in the main emission phase \citep{Chenjunping24}. 
    
\begin{figure*}
\centering
\begin{tabular}{cc}
\begin{overpic}[width=0.45\textwidth]{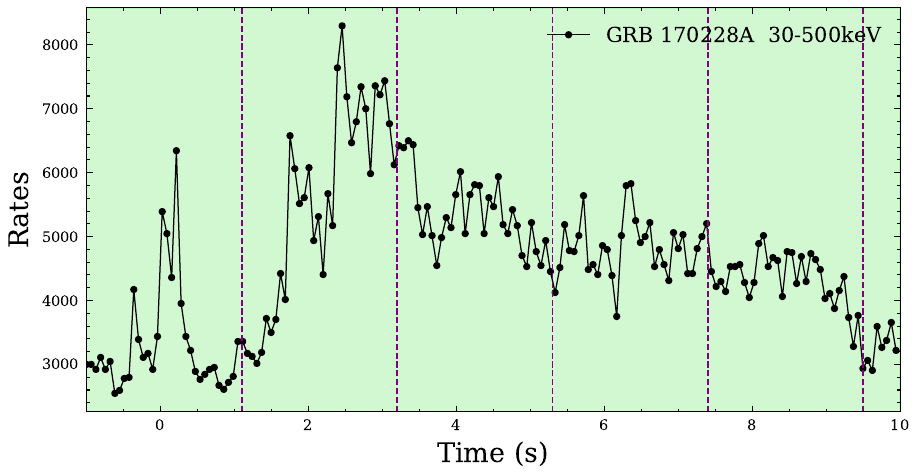}\put(0,55){\bf a}\end{overpic} &
        \begin{overpic}[width=0.45\textwidth]{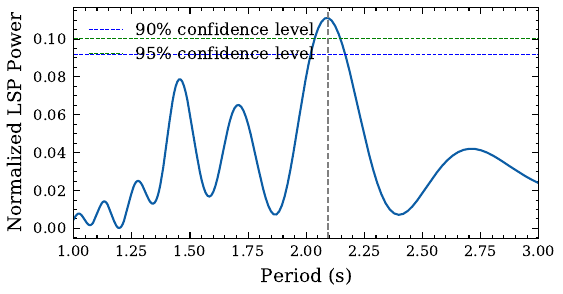}\put(0, 55){\bf b}\end{overpic} \\
\begin{overpic}[width=0.45\textwidth]{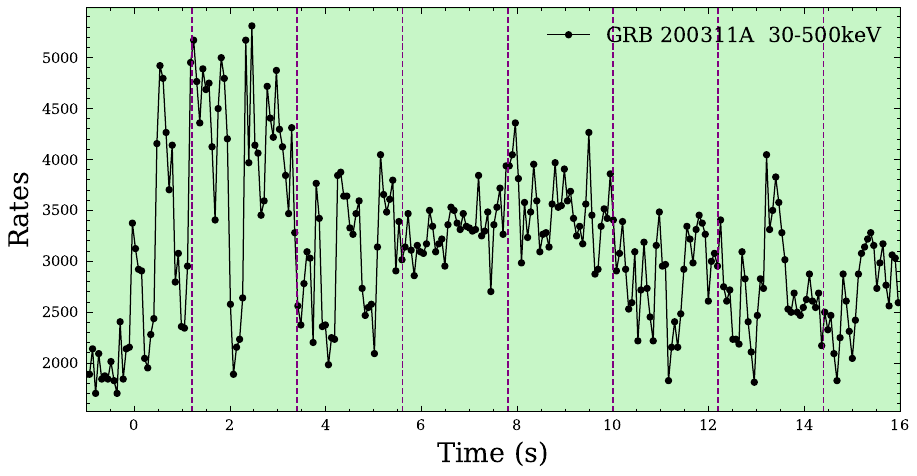}\put(0, 55){\bf c}\end{overpic} &
        \begin{overpic}[width=0.45\textwidth]{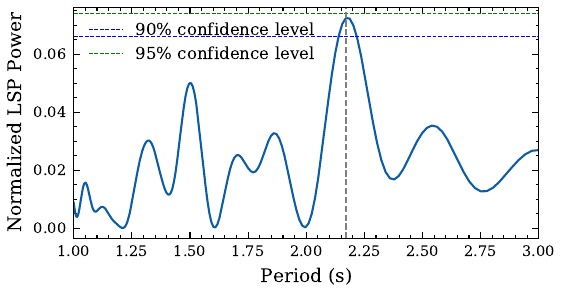}\put(0, 55){\bf d}\end{overpic} \\
\end{tabular}
\caption{\noindent\textbf{The QPM of GRB 170228A and 200311A.} 
\textbf{a-b}, the left panel is light curve of GRB 170228A in the 30–500 keV energy range, where the QPM episode is highlight by light green region. The right panel is the LSP power of the light curve.
\textbf{c-d}, the left panel is light curve of GRB 200311A in the 30–500 keV energy range, where the QPM episode is highlight by light green region. The right panel is the LSP power of the light curve.}
\label{fig:QPM}
\end{figure*}

    We perform the Lomb-Scargle Periodogram (LSP) \citep{LSP1976,LSP2018} to search for this periodicity in the remaining three Type IL GRBs. It turns out that during the main emission of GRB 170228A and GRB 200311A, the weak periodicity shows in the light curve, with the period about 2.1\,s and 2.2\,s, respectively (Fig.~\ref{fig:QPM}). The weak periodicity lasts for about 11\,s for GRB 170228A and 17\,s for GRB 200311A, corresponding to the $\sim 5$ cycles and $\sim 8$ cycles, respectively. Substituting the period we obtain in Equation (3) and using Equation (1) in \cite{Chenjunping24}, the mass of the WD is estimated to be $M_{\ast }  \sim  1.4 M_{\odot } $ for GRB 170228A, and GRB 200311A (taking $e\simeq 0$ ). The resulting WD mass is very close to the Chandrasekhar limit, which is consistent with the statements made in \cite{YJ_GRB211211A_nature}. This result further validates the scenario that the progenitor is the compact binary merger involving a WD, and the merger product (central engine) is a magnetar.

    It is worth noting that there could be another situation in which the duration of GRB can be explained in terms of the emitter rather than the progenitor or the center engine. That is, if the central engine impulsively shines while the jet itself continues to radiate through the dissipation process, then the observed duration of radiation is attributed to jet's own activity but not the progenitor or central engine's activity. Such a light curve would have an overall broad pulse profile, with the property of ``softer wider, softer later". In this case, we cannot determine the duration of the central engine activity, so neither NS-NS merger nor WD-NS merger can be ruled out. The typical model is the internal collision-induced magnetic reconnection and turbulence (ICMART) model, as used to explain GRB 230307A \citep{Yi07A2023}.

    \subsection{Extended emission}

    If the main emission is explained by the accretion of a magnetar, then the extended emission can be explained by subsequent magnetar activity.
    
\begin{figure*}
\centering
\begin{tabular}{cc}
\begin{overpic}[width=0.45\textwidth]{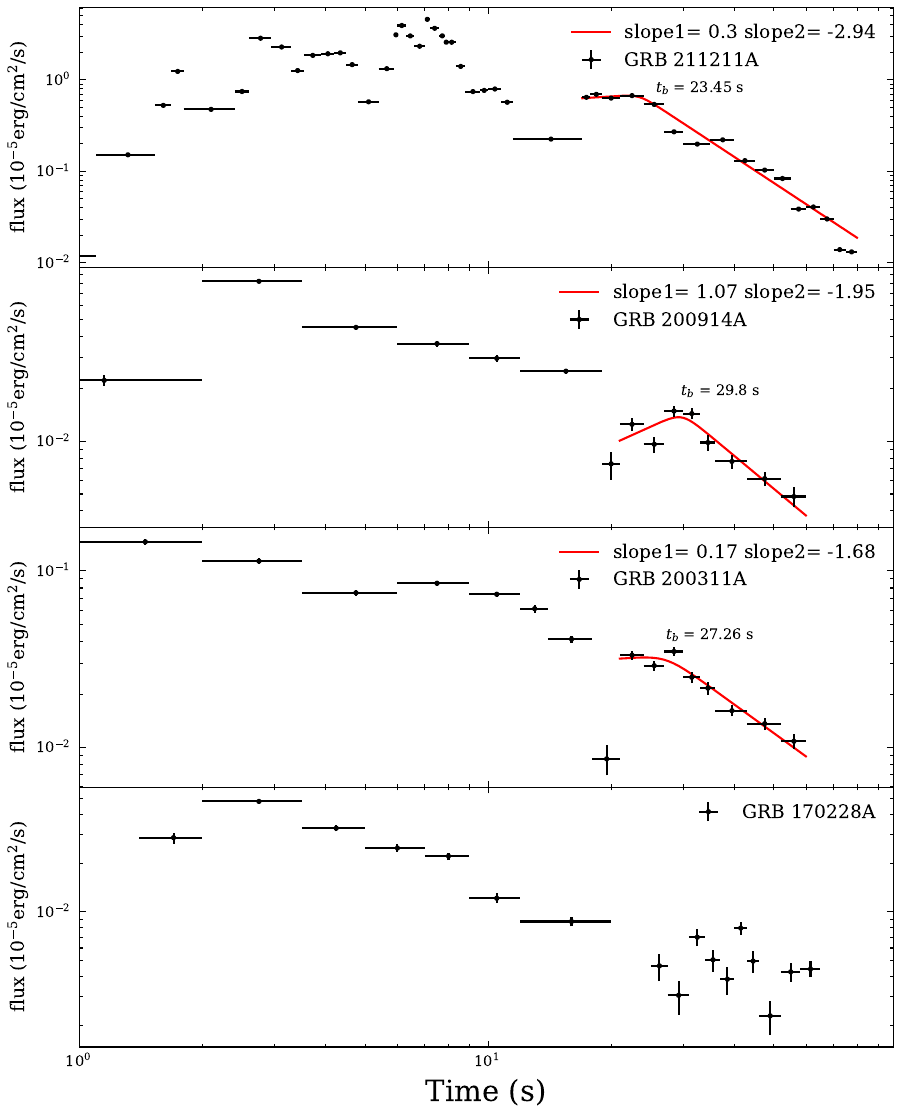}\put(0, 100){\bf a}\end{overpic} &
        \begin{overpic}[width=0.45\textwidth]{ 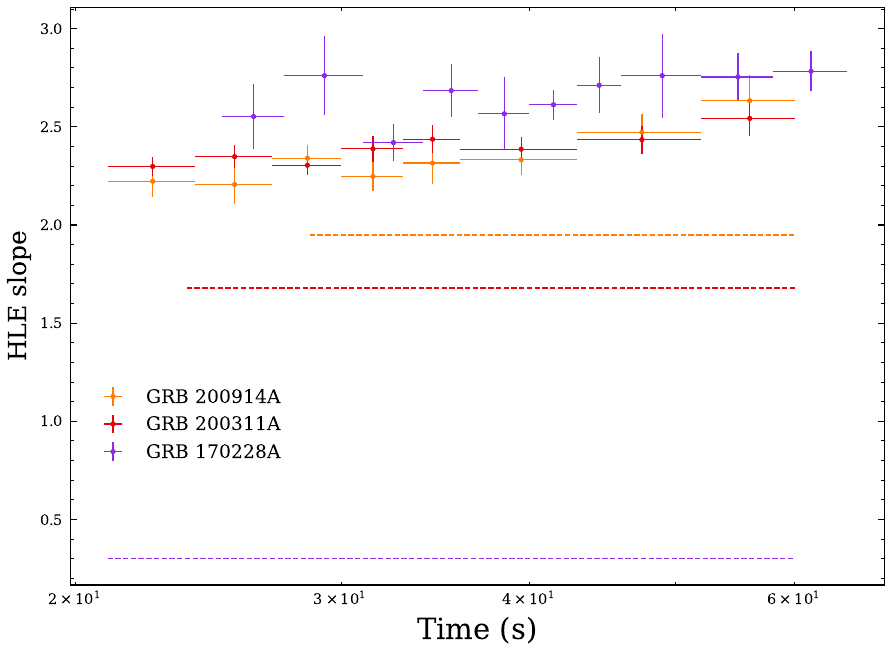}\put(0, 80){\bf b}\end{overpic} \\
\end{tabular}
\caption{\noindent\textbf{The flux evolution of four GRBs and the curvature effect daigram.} 
\textbf{a}, The flux of main emission and extended emission of four GRBs. The extended emission are fitted by the SBPL model (red line).
\textbf{b}, the dashed lines represent the HLE slope index calculated by the flux slope and the dots represent spectral index of extended emission. }
\label{fig:EEflux&HLE}
\end{figure*}

    We perform the time-resolved spectrum fitting of the extended emission of four Type IL GRBs, using the CPL model for GRB 211211A and the PL model for GRB 170228A, GRB 200914A and GRB 200311A due to the low statistic, and obtain the flux light curve. Then we perform an empirical fit to the extended emission flux light curve with a smoothed broken power-law(SBPL) model,

    \begin{equation}    
    F=(F_{1}^{-\omega }+F_{2}^{-\omega}   )^{-1/\omega },
    \label{equ:SBPL}
    \end{equation}
    
    where $F_{1}=A(t/t_{b})^{\alpha _{1} }$, $F_{2}=A(t/t_{b})^{\alpha _{2} }$. The power law slopes before and after break time $t_{b}$ are $\alpha _{1}$ and $\alpha _{2}$, respectively, and A is the normalization coefficient at $t_{b}$. $\omega$ describes the sharpness of the break at $t_{b}$, and is fixed to 10 in our fits. The fit of flux light curves is shown in Fig.~\ref{fig:EEflux&HLE}a. Overall, the extended emission of Type IL GRB exhibits the power-law decay following a flat plateau. The exception comes from GRB 170228A, which does not show the power-law decay after the plateau, probably because the intense of the decay stage is below the detection limit, as its flux is lower than other three GRBs in Fig.~\ref{fig:EEflux&HLE}a.

    Except for the extended emission of GRB 230307A, which is explained as a high-latitude effect, the extended emission of GRB 211211A was explained by the differential-rotation-induced magnetic bubbles model \citep{YJ_GRB211211A_nature}, which may also be suitable for the remaining three Type IL GRBs. However, here we still explore the possibilities for other models.

    The magnetar spin-down model has been widely invoked to interpret extended emission, and it seems to coherently explain the extended emission of Type IL GRBs \citep{2025NSRSun, houjun200219}. The magnetar loses rotational energy through electromagnetic radiation (magnetic dipole radiation), with the luminosity of the extended emission showing a plateau followed by a power-law decay characteristic with a slope of -2 \citep{2018GRBbook, Laskyspindown2016, houjunspindown2018}. 

    \begin{equation}    
     L = \left\{ 
    \begin{matrix}
    L_{0} , \qquad \qquad \qquad t\ll\tau _{c,em} \\
    L_{0}\left ( 1+\frac{t}{\tau _{c,em}}   \right ) ^{-2}, t\gg \tau _{c,em}\\
    \end{matrix}
    \right.
    \label{equ:spin-down}
    \end{equation}

    where $L_{0}$ is the initial luminosity,

    \begin{equation}    
    L_{0} = (1.0\times 10^{49}\ erg\ s^{-1}  )B_{p,15}^{2}P_{0,-3}^{-4}R_{6}^{6} 
    \label{equ:L0}
    \end{equation}

    $\tau _{c,em}$ is the characteristic spin-down timescales for electromagnetic radiation,

    \begin{equation}    
    \tau _{c,em} = (2.1\times 10^{3}\ s)I_{45} B_{p,15}^{-2}P_{0,-3}^{2}R_{6}^{-6} 
    \label{equ:tau}
    \end{equation}

    Where $I$ is the moment of inertia, $B_p$, $P_0$ and $R$ are the rotating period, surface magnetic field and radius of the neutron star, respectively. The convention $Q=10^xQ_x$ is adopted in cgs units.
    
     The model describes the data very well, with the only potential problem being that the break time is somewhat small, resulting a small $\tau _{c,em}$, which requires the magnetar to have either a smaller period or a stronger magnetic field. However, the energy range of the flux in our calculation might be too high, and the break time would be delayed if it were in the Swift/XRT energy range, resulting in the parameter falling within a reasonable space.

     In addition to the magnetic dipole radiation, another model invokes the relativistic stellar wind, which extracts rotational energy from the original magnetar \citep{Metzger08}. But this kind of model exhibits extended emission of hump-shaped, which is not consistent with power-law decay behavior. 

     Additionally, some models consider that the extended emission is generated by the interaction between the magnetosphere and the accretion disk, rather than the radiation from the magnetar itself \citep{GompertzEE14,Gibson17}. Their central engine invokes a magnetar surrounded by a fallback accretion disk, either formed from the merger of two compact objects or from a WD's accretion-induced collapse. During the extended emission phase, material is accelerated to super-Keplarian velocities and ejected from the system by the rapidly rotating and very strong magnetic field in a process known as magnetic propellering. The total luminosity consists of contributions from both propeller luminosity and magnetic dipole radiation luminosity. The disk is formed by the fallback accretion. During the coalescence process, some material is expelled due to tidal disruption and then re-falls for accretion assuming the accretion disk appears at t = 0 $s$, which means that accretion starts immediately and is at its peak intensity. In reality, material is still in the fallback period initially, thus, accretion is initially much gentler; however, as the accretion increases, the intensity also grows. \cite{Lee2009} predict that the material will return on a 10-second timescale. This seems to help explain the ~10\,s main emission of Type IL GRBs. However, the extended emission is also platform-like or hump-shaped for this model, and the main emission and the extended emission are produced by different processes in different regions, which contradicts the requirements mentioned earlier in Section \ref{section5}.
    
     In addition to invoking magnetar to explain the central engine and extended emission, a BH as the central engine has been established to depict the unified picture of merger type GRBs \citep{GottliebMAD23}. In this model, the BH is produced by compact binary mergers. The main emission generated by the Blandford-Znajek (BZ) mechanism and the magnetically arrested disk (MAD) is introduced to explain the extended emission. The power of the jet during the main emission is determined by the magnetic flux: the greater the magnetic flux, the higher the power. When the magnetic flux accumulates to saturation, the accretion disk enters the MAD phase, at which point the jet power is determined by the accretion rate. After entering the MAD phase, the accretion rate decreases over time following a power-law with slope of -2. Consequently, the time it takes for the accretion disk to enter the MAD phase determines the duration of the main emission. A larger accretion disk mass results in a later transition to MAD, leading to longer main emission durations. This model seems to coherently unify long-duration and short-duration merger bursts, and the power-law decay with slope of -2 in the extended emission also matches observational data. Moreover, both the main emission and the extended emission are produced by the same process, satisfying the requirements mentioned in Section \ref{section5}. Therefore, when discussing the central engine of Type IL GRB, this kind of scenario should not be excluded.

\section{Summary}\label{section6}
    Type IL GRB is an important subclass of GRBs. In this work, we establish the criterion of searching for Type IL GRBs. By searching in the \textit{Fermi GBM Burst Catalog}, we find five Type IL GRBs in the GRB samples from the past decade (i.e., 2014 to 2024). We successfully verify through their energy spectrum and temporal properties that they belong to a distinct subclass of Type I GRBs, clearly differentiating them from traditional ``long" sGRBs. The compact binary merger origin and long duration of the main burst make these GRBs ``special in special”. 
    
    In addition to the light curve patterns constrained by the criterion, we find that distribution of the duration of main emission in these 5 GRBs is relatively concentrated, and the energy ratio between extended emission and main emission is almost constant ($\sim0.7$, with small scattering), such commonality may have strong implication on the mechanism of Type IL bursts. We also search for thermal components in the precursors of these five Type IL GRBs and discuss the progenitor model and central engine model based on the long durations of the main emission, as well as briefly discussing the extended emission model. The existence of thermal component in precursor, the slope evolution behavior of extended emission, as well as the existence of weak periodic activity in the light curve, all indicate the diversity of Type IL GRB.
    
    Due to the lack of follow-up data of the X-ray range, we are unable to place strong constraints on the extended emission of Type IL GRB. However, this is precisely the significance of our work: to establish a standardized criterion for rapid recognition of this kind of distinctive GRB, thus guiding further low-latency follow-up observations, accumulating detailed data, and expanding our understanding of ``special in special” GRBs.

\section*{Acknowledgments}

We are grateful to Ruo-Yu Guan for useful discussions. We appreciate the anonymous reviewer for the helpful comments and suggestions. This work is supported by the National Natural Science Foundation of China (Grant No. 12494572, 12273042), the Strategic Priority Research Program of the Chinese Academy of Sciences (Grant No. XDA30050000, XDB0550300) and the National Key R\&D Program of China (2021YFA0718500). We appreciate the public data and software of \textit{Fermi}/GBM. The GECAM (Huairou-1) mission is supported by the Strategic Priority Research Program on Space Science (Grant No. XDA15360000) of Chinese Academy of Sciences.

\clearpage
\clearpage

\bibliography{reference}
\bibliographystyle{aasjournal}

\end{document}